\documentclass[%
reprint,
superscriptaddress,
 amsmath,amssymb,
 aps,
floatfix,
]{revtex4-1}

\usepackage[dvipsnames,svgnames,x11names,hyperref]{xcolor}
\usepackage{siunitx}
\usepackage{graphicx} 
\usepackage{amsmath}
\usepackage{hyperref}
\usepackage{multirow}
\usepackage{xcolor} %\usepackage{soul}
\usepackage{hyperref}
\usepackage{float}
\hypersetup{colorlinks, 
    linkcolor={blue!75!black!80!yellow},
    citecolor={blue!75!black!80!yellow}, 
    urlcolor={blue!75!black!80!yellow}
    }
\usepackage{graphicx}% Include figure files
\usepackage{dcolumn}% Align table columns on decimal point
\usepackage{bm}% bold math
\makeatletter \renewcommand\@make@capt@title[2]{%
\@ifx@empty\float@link{\@firstofone}{\expandafter\href\expandafter{\float@link}}%
\sffamily{\textbf{#1}}\@caption@fignum@sep#2 }% \makeatother
\newcommand{\HarvardSEAS}{John A. Paulson School of Engineering and Applied Sciences, Harvard University, Cambridge, MA 02138, USA}
\newcommand{\Sandia}{Sandia National Laboratories, Albuquerque, NM, USA}
\newcommand{\MIT}{Massachusetts Institute of Technology, Cambridge, MA, USA}
\newcommand{\RLE}{Research Laboratory of Electronics, Massachusetts Institute of Technology, Cambridge, MA, USA}

\begin{document} 
%\title{Coherent Interfaces for Hybrid \\Superconductor-Spin-Photon Quantum Information Processors}
%\title{Coherent Interfaces for Hybrid \\Superconductor-Spin-Photon Quantum Information Processors}

%Enabled by Quantum Transduction

%A Phononic Bus for Coherent Interfaces Between a Superconducting Quantum Processor, Spin Memory, and Photonic Quantum Networks

\title{A Phononic Bus for Coherent Interfaces Between a Superconducting Quantum Processor, Spin Memory, and Photonic Quantum Networks}

%Coherent Interfaces for Hybrid Superconductor-Artificial Atom\\ Quantum Information Processors with High-Fidelity Quantum Network Ports}

\author{Tom\'{a}\v{s} Neuman$^{+}$}
%\email{tomasneuman@seas.harvard.edu} 
\affiliation{\HarvardSEAS}

\author{Matt Eichenfield$^{+}$}
%\email{meichen@sandia.gov}
\affiliation{\Sandia}

\author{Matthew Trusheim$^{+}$}
\affiliation{\HarvardSEAS}\affiliation{\MIT}\affiliation{CCDC Army Research Laboratory, Adelphi, MD 20783, USA}

\author{Lisa Hackett}
\affiliation{\Sandia}

\author{Prineha Narang}
\email{prineha@seas.harvard.edu}
\affiliation{\HarvardSEAS}

\author{Dirk Englund}
\email{englund@mit.edu} 
\affiliation{\MIT}\affiliation{\RLE}

\begin{abstract}
% 
%Superconducting circuits have emerged as leading quantum computing platforms due to high-fidelity and high-speed qubit initialization and logic gates. Among the most important challenges now are to increase qubit coherence time, connectivity, and qubit numbers. To address these challenges, we introduce here a mechanism for high-speed, high-fidelity transduction between a qubit in a superconducting microwave circuit and diamond color center spin. 

%high-speed and high-fidelity single- and multi-qubit gates

%Superconducting (SC) qubits have been widely used as building blocks in quantum information technologies, including state-of-the-art quantum computers.

 %Linking quantum information processors requires the ability to interface the SC platform with travelling photons, and for long-distance quantum networking, to store quantum states for extended period of time in a quantum random-access memory (QRAM). 
 
% In this \emph{Article} we establish a scheme based on a high-quality acoustic resonator that enables high-speed and high-fidelity quantum-state transfer from superconducting qubits to electron-spin qubits in solid-state quantum defects, and finally to long-lived nuclear-spin qubits. 
 
We introduce a method for high-fidelity quantum state transduction between a superconducting microwave qubit and the ground state spin system of a solid-state artificial atom, mediated via an acoustic bus connected by piezoelectric transducers. Applied to present-day experimental parameters for superconducting circuit qubits and diamond silicon vacancy centers in an optimized phononic cavity, we estimate quantum state transduction with fidelity exceeding 99\% at a MHz-scale bandwidth. By combining the complementary strengths of superconducting circuit quantum computing and artificial atoms, the  hybrid architecture provides high-fidelity qubit gates with long-lived quantum memory, high-fidelity measurement, large qubit number, reconfigurable qubit connectivity, and high-fidelity state and gate teleportation through optical quantum networks. 

%mediated by piezoelectric actuation through a phononic bus. By combining the specific advantages of each qubit modality, the hybrid 

%photon architecture based on coherent transduction of quantum information between a superconducting quantum 

%scheme based on a high-quality acoustic resonator that enables high-speed and high-fidelity quantum-state transfer from superconducting qubits to electron-spin qubits in solid-state quantum defects, and finally to long-lived nuclear-spin qubits. 

%The fast and efficient optical addressability of the electron-spin allows for connectivity of the device to quantum networks via heralded entanglement generation, while the nuclear spins can store quantum information as a scalable QM. Successful state transfer is conditioned on optimal design of the acoustic cavity which efficiently couples the defect electron spin and the SC qubit with rates exceeding intrinsic system decay and decoherence. We show that high-fidelity phonon-mediated state transfer between a SC qubit and an electron spin can be achieved in realistic phononic cavities, and provide interconnects for integrating QM in large-scale quantum networks and quantum information processing.  

\end{abstract}

\date{\today}

\maketitle

\section{Introduction} 
%Coherent transduction of quantum states between disparate qubit modalities is an essential component in quantum information science. While superconducting circuits  have emerged as a leading platform with high-fidelity and high-speed qubit initialization and logic gates, there is a need for increased qubit coherence time, connectivity, qubit counts, and readout fidelity. 

% add citations Tomas/Matt?

%\cmmb{T: where should we cite optomechanics?}

Hybrid quantum systems have the potential to optimally combine the unique advantages of disparate physical qubits. In particular, while superconducting (SC) circuits have high-fidelity and high-speed  initialization and logic gates \cite{neeley2010generation, pop2014coherent,Ofek2016, narla2016entanglement,Lu2017-uc, barends2019diabaticgates, Arute2019-cd, Kjaergaard_2019}, challenges remain in improving qubit (i) coherence times, (ii) long-range connectivity, (iii) qubit number, and (iv) readout fidelity. A hybrid system may satisfy these challenges by delegating different tasks to constituent physical platforms. Here, we propose an approach to enable such scalable solid-state quantum computing platforms, based fundamentally on a mechanism for high-fidelity qubit transduction between a SC circuit and a solid-state artificial atom (AA). 
Mediating this transduction is an acoustic bus \cite{Chu2017-wl,kuzyk2018scaling, li2019honeycomb, Bienfait2019transferSAW} that couples to the SC qubit and an AA electron spin via a combination of piezoelectric transduction and strong spin-strain coupling. Applied to present-day experimental parameters for SC flux qubits and silicon vacancy (SiV$^-$) centers in diamond, we estimate quantum state transfer with fidelity exceeding 99\% at a MHz-scale bandwidth. Hyperfine coupling to local ${^{13}}$C nuclear-spin qubits enables coherence times exceeding a minute \cite{bradley2019register}, while excited orbital states enable long-distance state transfer across quantum networks by optically heralded entanglement. Moreover, the scheme is extensible to large numbers of spin qubits with deterministic addressability, potentially enabling integration of large-scale quantum memory. Noting that SiV$^-$ single-shot optical readout fidelity has been experimentally demonstrated to exceed 99.9\% \cite{Bhaskar2019-gd}, this approach thus successfully addresses challenges (i-iv). By combining the complementary strengths of SC circuit quantum computing and artificial atoms, this hybrid SC-AA architecture has the essential elements for extensible quantum information processors: a high-fidelity quantum processing unit (QPU), a bus to scalable quantum memory, and a high-fidelity connection long-range optical quantum networks.

\begin{figure*}
    \centering
    \includegraphics[scale=0.39]{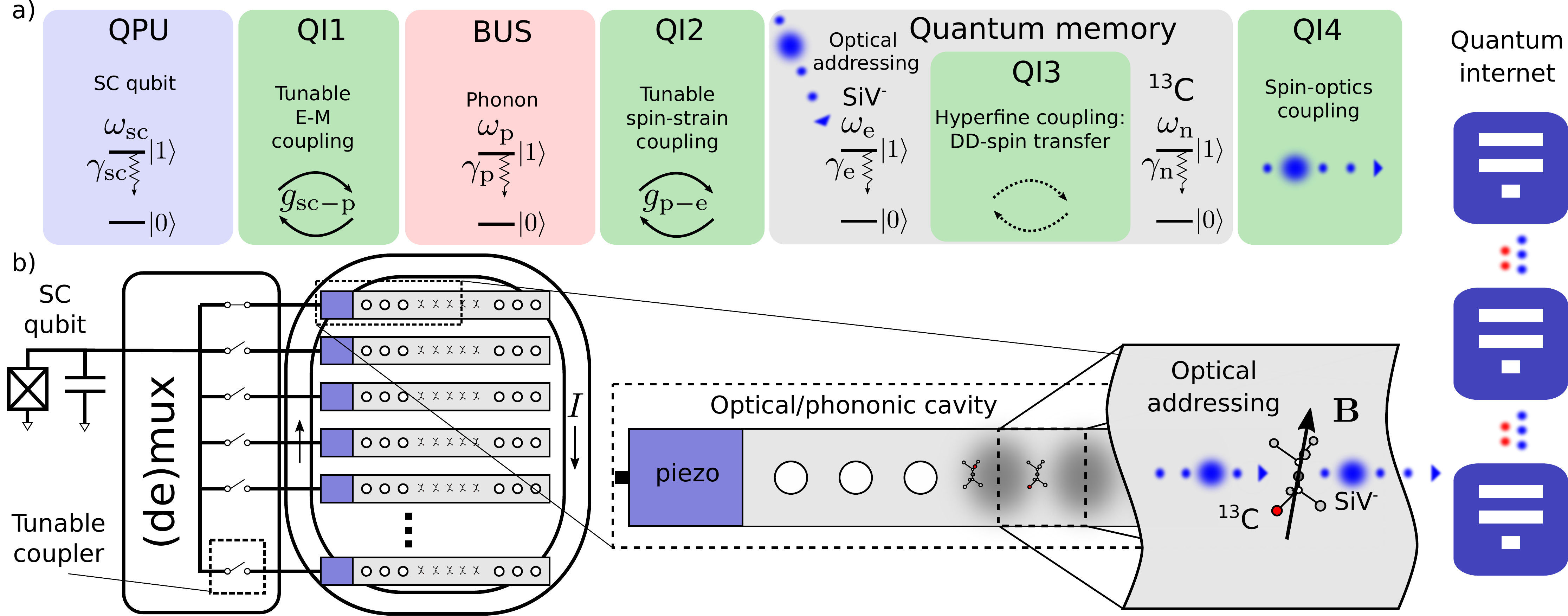}
    \caption{Quantum Memory (QM) and Interconnect Architecture
    (a) A SC quantum processing unit (QPU) is connected via piezoelectric `quantum interface 1' (QI1) to a phononic BUS. The phonon couples to electronic spin-orbit states of an AA, forming quantum interface 2 (QI2). The AA's fine-structure states can further couple to nuclear-spin to realize a QM via quantum interface 3 (QI3) or they connect to photons in quantum interface 4 (QI4), which finally connects to the quantum internet (blue dots: photonic interconnects). 
    (b) A physical realization of the scheme outlined in (a). A superconducting qubit is connected via a phononic or microwave multiplexer (mux) to a series of phononic or microwave waveguides that are each interfaced with a mechanical cavity hosting one or many AAs whose electronic fine-structure (spin-orbit) states serve as qubits. 
    The spin-orbit states of each AA interacts with the spin states of a nearby $^{13}$C nucleus via the hyperfine interaction, providing ancillae qubits with long coherence times. The AA optical transitions provide an optical interface to quantum networks, for example by multiple attempts of photon-based electron spin entanglement to provide an on-demand entanglement resource. Underlying these capabilities is a transduction scheme that exploits: tunable electro-mechanical (E-M) coupling between the SC qubit and the mechanical mode, tunable spin-strain coupling between the mechanical mode and the electron spin, optical addressing of the electron spin, and hyperfine coupling to connect the electron spin to the nuclear QM.}
    \label{fig:schematicfig1}
\end{figure*}

Our approach, schematically depicted in Fig.\,\ref{fig:schematicfig1}, combines four quantum interfaces [QIs, marked as QI1-QI4 in Fig.\,\ref{fig:schematicfig1}(a)] between physical modalities: a microwave photon-to-phonon interface, coupling of a phonon to an AA electron spin, coupling of the electron spin to a nuclear spin, and finally coupling of the electron spin to the optical photon. Previous work has investigated these quantum interfaces separately, including the piezoelectric transduction from the microwave circuit to the phonon~\cite{schuetz2015universaltransducers, Manenti2017, arrangoiz2018superconducting, Bienfait2019transferSAW, hann2019hardwareefficient,  Higginbotham2018harnessing, sletten2019phononfock, wu2020microwave}, spin-strain coupling in solid-state quantum emitters \cite{falk2014electrically, golter2016optomechNV, kuzyk2018scaling, lemonde2018phononnetworks,chen2018orbital, maity2018alignment, meesala2018strainsiv, udvarheliy2018spinstrain, li2019honeycomb}, hyperfine interactions of electron spins with nearby nuclei \cite{de2010universal, Childress281, taminiau2014universal,Waldherr2014, bradley2019register, nguyen2019nuclearoptics}, and spin-dependent optical transitions~\cite{Pfaff532, Bernien2013, Evans2018-vh,Awschalom2018-en}.
The last mentioned, optical response of AAs conditioned on the electron spin state, can be used to generate heralded entanglement~\cite{humphreys2018deterministic, rozpkedek2019near, bhaskar2019experimental} and thus allow for networking (e.g. connecting the device to the quantum internet) via quantum-state teleportation. As compared to optomechanical~\cite{stannigel2010transducer,stannigel2011optomechanicaltransducer, Bochmann2013} and electro-optical~\cite{Rueda2016} transduction schemes, quantum teleportation circumvents the direct conversion of quantum states into photons and thus minimizes the infidelity associated with undetected (unheralded) photon loss. 
Recent experiments have demonstrated the strain-mediated driving of an AA electron spin ground state with a classical phonon field~\cite{Whiteley2019-xe, Maity2020straincontrol}. Using the strain-spin coupling rates measured in those experiments to inform a theoretical model, and introducing a new phononic cavity design that achieves the strong coupling regime between a single phonon and an AA spin, we estimate that quantum state transduction is possible with near-unity fidelity, as shown below. 
%%% TOMAS END

%Coherent transduction has been experimentally demonstrated for all other QIs. 

%Notably, optomechanical~\cite{stannigel2010transducer,stannigel2011optomechanicaltransducer, Bochmann2013} and electro-optical~\cite{Rueda2016} transduction schemes interfacing SC quantum circuits with traveling photons have been proposed. These schemes rely on efficient emission and absorption of photons. In contrast, our proposed scheme takes advantage of quantum-state teleportation, circumventing this direct transduction of solid-state qubits to and from photons. 
%by employing heralded quantum teleportation protocols exploiting the deterministically prepared bell states shared between quantum nodes. 

%In the following, analyze the performance of the proposed SC-AA architectured based on realistic experimental parameters to demonstrate potential of this architecture for scaling quantum information processing.

The \textit{Article} is structured as follows. Section~\ref{sec:II} develops a general model for phonon-to-spin transduction using a quantum master-equation approach, followed in Section~\ref{sec:III} by experimentally informed model parameters. Section~\ref{sec:IV} introduces designs for mechanical cavities that achieve strong phonon-spin coupling and efficient quantum state transfer through a combination of high zero-point strain amplitude at AA sites and high expected mechanical quality factors. In Section~\ref{sec:V}, we numerically evaluate the master equation describing SC-electron spin transfer and demonstrate a state transfer infidelity below $\sim 1$\%, and even below 0.1\% (sufficient for fault tolerance threshold) using more speculative techniques. In Section~\ref{sec:VI}, we elaborate using the AA's optical transitions to realize optical interconnects and -- by heralded entanglement to other networked quantum memories -- to enable on-demand, long-range state and gate teleportation with near-unity fidelity.

\section{Theoretical model of the quantum-state transduction}\label{sec:II}

To estimate the state-transfer fidelity we theoretically model the quantum state transfer from the SC qubit to the electron-spin qubit using the quantum-master-equation approach. As in Fig.\,\ref{fig:schematicfig1}, the SC qubit is directly coupled to a discrete mechanical mode of a phononic cavity via a tunable electromechanical transducer. In Appendix\,\ref{app:indirect} we describe an alternative coupling scheme in which the interaction between the SC qubit and the mechanical mode of the cavity is mediated by guided modes of a microwave \cite{wu2020microwave} or phononic waveguide \cite{Fang2016transductionphonons, Bienfait2019transferSAW}. These guided modes mediate the state transfer between the SC qubit and the discrete phononic mode. The couplings to and out of the waveguide are time-modulated to release (``pitch'') and later catch a wavepacket of propagating waveguide modes. Finally, the strain of the mechanical mode interacts with spin levels of the electronic fine-structure states of a diamond AA. By controlling this coupling, the quantum state is transduced to the spin state of the AA electron. 

We start our theoretical description from the Hamiltonian schematically depicted in Fig.\,\ref{fig:schematicfig1}(a):
\begin{align}
    H_{\rm sc-e}&=\hbar\omega_{\rm sc}\sigma_{\rm sc}^\dagger \sigma_{\rm sc} + \hbar \omega_{\rm p} b^\dagger b + \hbar \omega_{\rm e}\sigma_{\rm e}^\dagger \sigma_{\rm e} \nonumber\\
    &+ \hbar g_{\rm sc-p}(t) (\sigma_{\rm sc}b^\dagger   + \sigma_{\rm sc}^\dagger b)\nonumber\\
    &+\hbar g_{\rm p-e}(t) (\sigma_{\rm e} b^\dagger + \sigma_{\rm e}^\dagger b).
\end{align}
Here $\sigma_{\rm sc}$ ($\sigma_{\rm sc}^\dagger$) is the superconducting qubit two-level lowering (raising) operator, $\sigma_{\rm e}$ ($\sigma_{\rm e}^\dagger$) is the electron spin lowering (raising) operator, and $b$ ($b^\dagger$) is the annihilation (creation) operator of the phonon. The frequencies $\omega_{\rm sc}$, $\omega_{\rm p}$, and $\omega_{\rm e}$ correspond to the SC, phonon, and electron-spin excitation, respectively. The SC couples to the phonon mode via the coupling rate $g_{\rm sc-p}$, and the phonon couples to the electron spin via $g_{\rm p-e}$. The operators $\sigma_{\rm sc}$ ($\sigma_{\rm sc}^\dagger$) describe the SC system in a two-level approximation and can be identified with the annihilation (creation) operators of the qubit flux appearing in the circuit cavity-QED description of the device \cite{blais2004circuitqed, devoret2004superconducting}. Throughout the paper we assume that all effective couplings in the system are resonant and thus $\omega_{\rm sc}=\omega_{\rm p}=\omega_{\rm e}$.

We consider system losses by adding into the Liouville equation of motion for the density matrix $\rho$ the Lindblad superoperators $\gamma_{c_i}\mathcal{L}_{c_i}(\rho)$:
\begin{align}
    \frac{\rm d}{{\rm d}t}\rho = \frac{1}{{\rm i}\hbar}[H_{\rm sc-e}, \rho]+\sum_i \gamma_{c_i}\mathcal{L}_{c_i}(\rho),\label{eq:mastereqdirect}
\end{align}
where 
\begin{align}
    \gamma_{c_i}\mathcal{L}_{c_i}(\rho)=\frac{\gamma_{c_i}}{2} \left(2 c_i\rho c_i^\dagger-\lbrace c_i^\dagger c_i,\rho \rbrace \right), \label{eq:lindblad}
\end{align}
with $c_i\in \{ \sigma_{\rm sc}, b,\sigma_{\rm e}^\dagger \sigma_{\rm e} \}$, and $\gamma_{c_i}\in\{ \gamma_{\rm sc}, \gamma_{\rm p}, \gamma_{\rm e}\}$ representing the decay (decoherence) rates of the respective excitations. We note that the Lindblad superoperators $\mathcal{L}_{\sigma_{\rm sc}}(\rho)$ and $\mathcal{L}_{b}(\rho)$ describe the $T_1$ processes including the qubit decay, whereas $\mathcal{L}_{\sigma^\dagger_{\rm e}\sigma_{\rm e}}(\rho)$ describes pure dephasing of the electron spin (a $T_2$ process) considering the long-lived character of the spin excitation. We do not include pure dephasing ($T_2$) processes of the SC qubit and the mechanical mode, but consider rates of the $T_1$ processes corresponding to the experimentally achievable $T_2$ times (since $T_1\sim T_2$ for phonons and SC qubits). We do not include thermal occupation of modes as we consider the system to be cooled to $\sim$mK temperatures.  

For high-fidelity state transfer without coherent reflections, it is necessary to switch the magnitude of the Jaynes-Cummings couplings $g_{\rm p-e}$ and $g_{\rm sc-p}$ in a sequence that allows for step-wise transfer of the quantum state to the mechanical mode and finally to the electron spin. To that end we first switch on the coupling $g_{\rm p-e}$ between the SC qubit and the mechanical mode while turning off the phonon-electron-spin coupling $g_{\rm sc-p}$. After completing the state transfer to the mechanical mode, we switch off $g_{\rm sc-p}$ and apply a state-transfer pulse $g_{\rm e-p}$ completing the procedure. Each of the pulses represents a SWAP gate (up to a local phase), so the state-transfer protocol can be inverted by interchanging the pulse order. In particular, we assume that each coupling has a smooth time dependence given by
\begin{align}
    g_{\rm sc-p}(t)&=g_{\rm scp}\,{\rm sech}(2 g_{\rm scp}[t-\tau_{\rm scp}])\label{eq:pulsescp}\\
    g_{\rm p-e}(t)&=g_{\rm pe}\,{\rm sech}(2 g_{\rm pe}[t-\tau_{\rm pe}]),\label{eq:pulsepe}
\end{align}
where $g_{\rm scp}$, $g_{\rm pe}$ are time-independent amplitudes and $\tau_{\rm scp}$, $\tau_{\rm pe}$ are time delays of the respective pulses. We choose the smoothly varying pulses over rectangular pulses to account for the bandwidth-limitation of experimentally achievable time-dependent couplings. In our simulations we adjust $\Delta\tau_{\rm sc-p-e}=\tau_{\rm pe}-\tau_{\rm scp}$ to optimize the state-transfer fidelity $\mathcal{F}$ defined as:
\begin{align}
    \mathcal{F}=\left| {\rm Tr} \left \{  \sqrt{\sqrt{\rho_{\rm i}}\rho_{\rm f}\sqrt{\rho_{\rm i}}} \right\}    \right|,\label{eq:fidelity}
\end{align}
where $\rho_{\rm i}$ ($\rho_{\rm f}$) is the density matrix of the initial state of the SC qubit (final state stored in the electron spin). Due to the finite simulation time we further approximate the ideal infinite time spread of the applied pulses and apply the pulses at a sufficient delay after the start of the simulation.  

\section{Physical Transducer Parameterization}\label{sec:III}
\subsection{SC Transducer Parameterization}\label{sec:III_A}

The values of the coupling and loss parameters govern the system performance. Coupling rates between a microwave (MW) resonator and a phononic cavity \cite{wu2020microwave} up to $\sim 100$ MHz have been shown experimentally in a MW-cavity resonantly coupled to a discrete phononic mode via an piezoelectric coupler. 
Optimizing the coupling requires matching the MW line impedance with the phonon waveguide impedance~\cite{siddiqui2018lambwave}. For tunable coupling between the SC qubit and the mechanical mode of the cavity, the MW resonator can be substituted by the SC qubit itself as in a recent experimental demonstration \cite{Bienfait2019transferSAW}. Using a Josephson junction with externally controllable flux as a tunable microwave switch mated   \cite{chen2014tunecoupler,geller2015tunable,zeuthen2018electrooptomech, Bienfait2019transferSAW} to the piezoelectric coupler thus enables controllable coupling between the SC qubit and the phonon.%, as illustrated in Fig.\,\ref{fig:controllablesc}.
%\begin{figure}
%    \centering
%    \includegraphics[scale=0.88]{Figuregscp_01.pdf}
%    \caption{Schematics of the SC qubit coupled to a mechanical mode of a phononic cavity. (a) The SC qubit behaves as a nonlinear MW resonator. (b) The mechanical mode is driven by an interdigital transducer (IDT) connected to the SC qubit via a MW circuit. (c) The coupling between the mechanical mode and the SC qubit can be controlled via an external magnetic flux applied to a Josephson junction \cite{chen2014tunecoupler,geller2015tunable,zeuthen2018electrooptomech, Bienfait2019transferSAW}.  }
%    \label{fig:controllablesc}
%\end{figure}
Based on recent experiments \cite{Bienfait2019transferSAW}, we assume that the coupling between the SC qubit and the mechanical mode can reach up to $g_{\rm scp}/(2\pi)=50$\,\si{\mega\hertz} . We conservatively assume SC qubit coherence times on the order of microseconds ($\gamma_{\rm sc}/(2\pi)=10$\,\si{\kilo\hertz}), while best-case SC coherence times approach milliseconds \cite{Kjaergaard_2019}.
%[cite yale Quantum memory with millisecond coherence in circuit QED]. 
\subsection{Spin-Strain Transducer Parameterization}\label{sec:III_B}

We consider that the spin qubit is formed by the two low-energy fine-structure states of the SiV$^-$ as described in Appendix\,\ref{app:strain}. These two states have distinct orbital and spin character which impedes direct coupling of the spin-qubit transition to either strain or magnetic fields. Generally, a combination of applied strain and magnetic field is thus necessary to address the SiV$^-$ spin qubit. We thus control the spin-strain coupling via locally applied magnetic fields to realize the effective controllable Jaynes-Cummings interaction introduced in Sec.\,\ref{sec:II} [Eq.\,\eqref{eq:pulsepe}]. Several strategies have been devised to engineer the effective spin-strain coupling \cite{meesala2018strainsiv, nguyen2019strainsi} that generally rely on the application of external static or oscillating magnetic fields and optical drives as we detail in Appendix\,\ref{app:strain}. All of these approaches are perturbative in character and the maximum achievable value of the resulting effective spin-strain coupling $g_{\rm pe}$ is therefore decreased with respect to the bare strain coupling measured for fine-structure spin-allowed transitions $g_{\rm orb}$ to $g_{\rm pe}\approx 0.1 g_{\rm orb}$. 
The spin-strain interaction of group-IV quantum emitters in diamond has been measured at 1\,\si{\peta\hertz}/strain \cite{nguyen2019strainsi}. We estimate that for an efficient state transfer between the mechanical mode and the electron-spin states, the spin-mechanical coupling $g_{\rm orb}$ would need to reach a value of approximately $g_{\rm orb}/(2\pi)\approx 10$\,\si{\mega\hertz} (leading to the effective phonon-electron-spin $g_{\rm pe}/(2\pi)\approx 1$~\si{\mega\hertz}). To that end a mechanical resonance with zero-point strain of $\sim 10^{-9}-10^{-8}$ and a high quality factor is needed. In Section\,\ref{sec:IV} we design (opto-)mechanical cavities that fulfill both of these requirements.

%[In an appropriately designed mechanical resonator featuring a small elastic mode volume, the single-phonon strain at the location of a diamond quantum emitter can reach $\sim 10^{-9}-10^{-8}$, leading to estimates of the single-phonon spin-allowed coupling rates of $\sim 10$\,\si{\mega\hertz}. REPHRASE to state desired parameters from phononic resonator] However, as we discuss in Appendix\,\ref{app:strain}, the transition between the spin-qubit states $|0\rangle$ and $|1\rangle$ is spin forbidden and the effective (tunable) strain coupling $g_{\rm p-e}$ to the spin-qubit is thus further reduced to $< 10$\% of the spin-allowed value $g_{\rm pe}\approx 0.1 g_{\rm orb}\approx 1$\,\si{\mega\hertz}.
%[Add actual value]. [[]even in the perturbative regime\ref{app:strain}??]
%\cmmb{T: No, this is the "bare coupling" ("spin-allowed"), the effective "spin forbidden" coupling will be smaller, we have to make this clear}].

\section{Realization of cavity for strong phonon-spin coupling}\label{sec:IV}
%Mechanical resonators operating at \si{\giga\hertz} frequencies with loss rates well below the \si{\kilo\hertz} range have been demonstrated, with heterogeneous piezoelectric integration.

This section introduces a mechanical cavity that allows fast and efficient phonon-mediated quantum-state transduction to and from the electron spin. We model the cavities through a series of finite-element numerical simulations \cite{Eichenfield2009} (performed using Comsol Multiphysics \cite{comsol}) of the mechanical resonance within the continuum description of elasticity. These simulations use absorbing perfectly matched layers at the boundaries. We obtain the optical response of the diamond cavity from a solution of Maxwell's equations in the materials described via their linear-response dielectric function. 

\begin{figure*}
    \centering
    \includegraphics[scale=0.35]{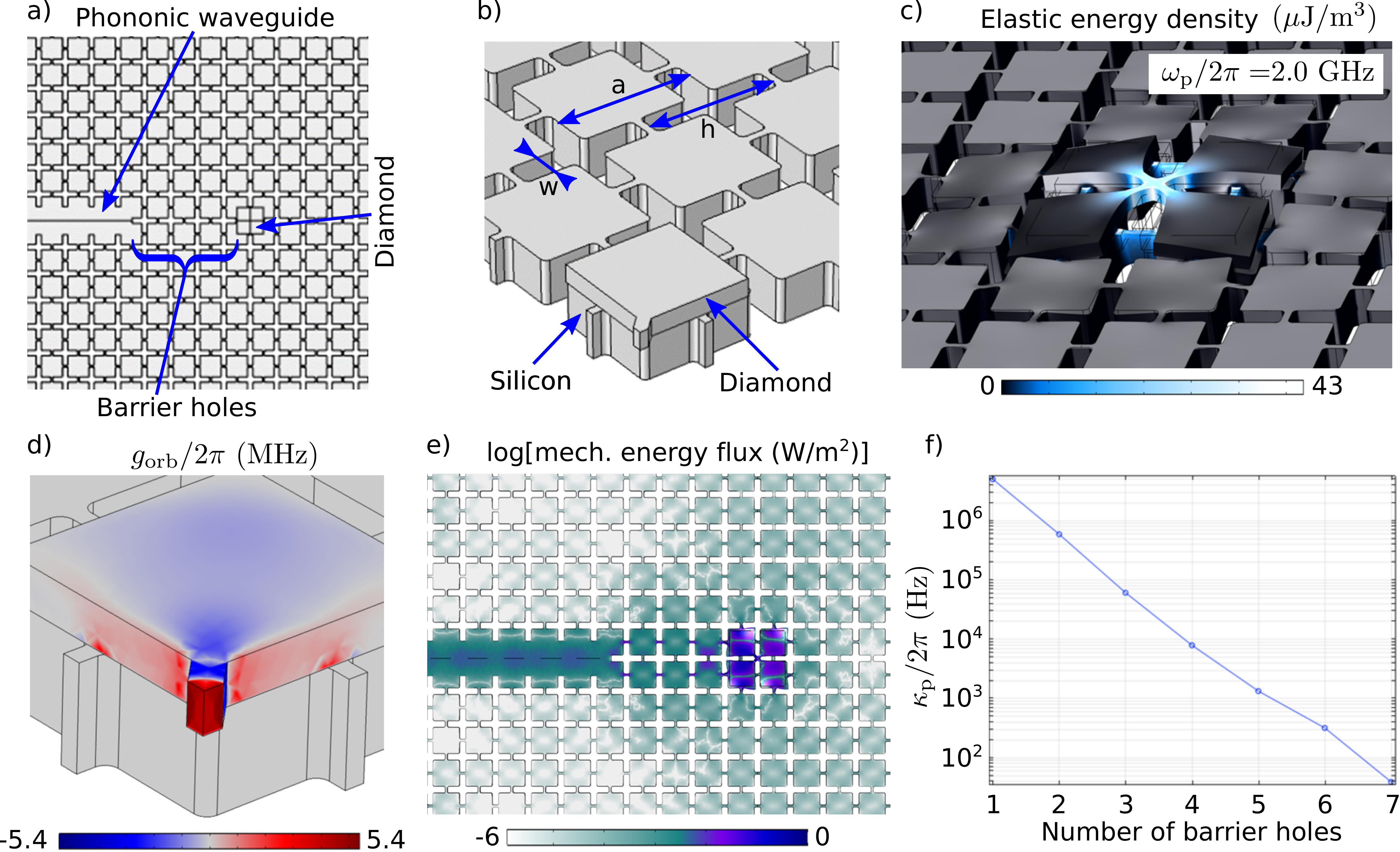}
    \caption{Silicon phononic cavity. (a) A mechanical resonator embedded in a phononic crystal is separated from a phononic waveguide by a number of barrier holes. It is capped by a thin diamond layer placed over the silicon layer. (b) A close-up of one quarter of the silicon-diamond structure with dimensions $a=800$\,nm, $h=0.94a$, and $w=0.2a$. The silicon (diamond) layer thickness is $t_{\rm Si}=250$\,nm ($t_{\rm D}=100$\,nm). The width of the thin rectangular diamond interconnects is $w_{\rm D}=70$\,\si{\nano\meter}. (c) The elastic energy density of the cavity mode is concentrated in the thin constriction of the diamond layer. The geometry of the cavity is artificially distorted along the coordinate of the mechanical mode. (d) The coupling rate between the mechanical mode and the electron spin calculated as $g_{\rm orb}\approx (\epsilon_{xx}-\epsilon_{yy})d$, with $d\approx 1$\,\si{\peta\hertz}/strain being the strain susceptibility of SiV$^-$ defect. (e) The mechanical energy flux of the combined eigenmode of the mechanical cavity and the waveguide. (f) Mechanical damping $\kappa_{\rm p}$ of the phononic cavity as a function of the number of barrier holes separating it from the phononic waveguide.}
    \label{fig:phononspin1}
\end{figure*}

\begin{figure}
    \centering
    \includegraphics[scale=0.37]{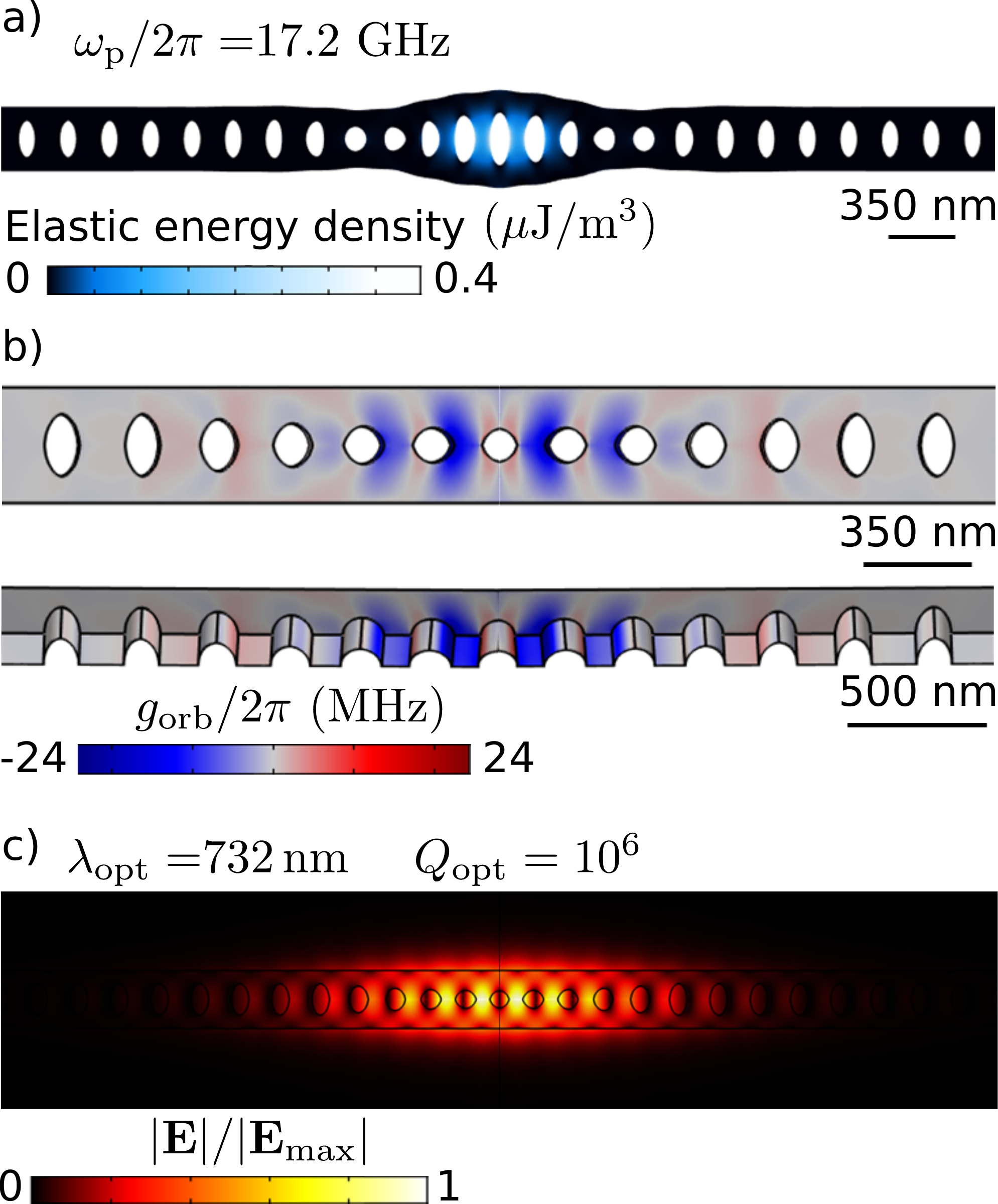}
    \caption{Diamond optomechanical cavity. (a) Distribution of elastic energy density of a mode of the diamond cavity resonant at $\omega_{\rm p}/(2\pi)=2$\,\si{\giga\hertz}. The cavity geometry is artificially distorted along the coordinate of the mechanical mode for emphasis. The energy density is concentrated in the central part of the cavity. (b) Calculated coupling rate $g_{\rm orb}$ in the phononic cavity. (c) Normalized amplitude of the optical electric field at vacuum wavelength $\lambda_{\rm opt}=732$\,\si{\nano\meter} of an optical mode of the diamond cavity of an optical quality factor $Q_{\rm opt}\approx 10^6$. The electric field is concentrated around the center of the cavity and thus overlaps with the regions providing high values of $g_{\rm orb}$. }
    \label{fig:phononspin2}
\end{figure}

We describe two architectures of high-$Q$ mechanical cavities whose zero-point strain field gives rise to the phonon-spin strong coupling required in the transduction scheme. (i) The first design is a silicon cavity with a thin (100 nm) layer of diamond heterogeneously integrated to the silicon substrate [shown in Fig.\,\ref{fig:phononspin1}]. This  takes advantage of mature design and fabrication of silicon nanophononics \cite{Eichenfield2009, Safavi-Naeini2011, Chan2011}, exceptionally small decoherence rates of microwave frequency phonons in suspended single crystal silicon \cite{maccabe2019phononic}, and new techniques in heterogeneous integration of diamond nanoscale membranes \cite{mouradian2015scalable, wan2019largescale}. (ii) The second design is an all-diamond optomechanical cavity that at the same time supports an optical and phononic mode for mechanical and optical addressing of the electron spin [shown in Fig.\,\ref{fig:phononspin2}]. As depicted in Fig.\,\ref{fig:phononspin1}(a), the silicon cavity is embedded in a phononic crystal to minimize the cavity loss; it is also weakly coupled to a phononic waveguide that mediates the interaction of the cavity with the SC circuit. Simultaneous acoustic and microwave electrical impedance matching has been demonstrated \cite{eichenfield2013design, siddiqui2018lambwave} to such wavelength-scale structures using thin piezoelectric films, enabling coupling into the waveguide from the superconducting system with low insertion loss. The cavity is separated from the waveguide by a series of barrier holes to allow tuning the coupling rate between the discrete cavity mode and the guided phonons. Fig.\,\ref{fig:phononspin1}(b) details this cavity. The silicon platform forming the base of the cavity is covered with a thin layer (100 nm) of diamond hosting the defects. 

To analyze the cavity mechanical properties, we calculate the distribution of the elastic energy density of a mechanical mode of the cavity resonant at $\omega_{\rm p}/(2\pi)=2.0$\,\si{\giga\hertz}. The energy density is concentrated in the thin constriction formed by the diamond layer for efficient phonon-spin coupling. Figure\,\ref{fig:phononspin1}(d) shows the calculated bare phonon-spin coupling $g_{\rm orb}=(\epsilon_{xx}-\epsilon_{yy})d$ corresponding to the strain field in the ground state of the resonator--the zero-point strain. Here $d\approx 1$\,\si{\peta\hertz}/strain is the strain susceptibility of the defect electron spin, and $\epsilon_{xx}$ ($\epsilon_{yy}$) are the components of the zero-point strain expressed in the coordinate system of the defect (see Appendix\,\ref{app:transformed}). As we show in Fig.\,\ref{fig:phononspin1}(d), $g_{\rm orb}/(2\pi)$ reaches up to $5.4$\,\si{\mega\hertz} and we thus estimate the effective coupling $g_{\rm pe}/(2\pi)\approx 0.5$\,\si{\mega\hertz}. An equally important figure of merit characterizing the cavity performance is the cavity coupling to the waveguide modes. The distribution of the mechanical energy flux in the cavity in Fig.\,\ref{fig:phononspin1}(e) shows that the cavity mode interacts with the waveguide modes, introducing a decay rate $\kappa_{\rm p}$ of the cavity mode. Fig.\,\ref{fig:phononspin1}(f) plots $\kappa_{\rm p}$ as a function of the number of barrier holes on a logarithmic scale. Thus, $\kappa_{\rm p}$ decreases exponentially with the number of separating barrier holes from almost $\sim 10^7$\,\si{\hertz} to $\sim 1$\,\si{\hertz} for seven holes. For larger number of holes, the cavity lifetime becomes practically limited by the intrinsic material properties of silicon and diamond and could be as low as $\sim 0.1$\,\si{\hertz} \cite{maccabe2019phononic} assuming no additional loss due to the introduction of the diamond nanomembrane.

Fig.\,\ref{fig:phononspin2} shows the all-diamond optomechanical cavity, which consists of a diamond beam with an array of elliptical holes of varying sizes. The hole array simultaneously produces a phononic and photonic cavity that concentrates both the mechanical strain of the phononic mode and the optical electric field of the electromagnetic mode on the AA in the cavity center. The distribution of the elastic energy density of a phononic mode of frequency $\omega_{\rm p}/(2\pi)=17.2$\,\si{\giga\hertz} shown in Fig.\,\ref{fig:phononspin2}(a) reveals that the mechanical energy is dominantly concentrated around the center of the beam. Using the calculated values of zero-point strain of this mode we calculate the achievable bare coupling strength $g_{\rm orb}$ and show the result in Fig.\,\ref{fig:phononspin2}(b). The maximal achievable effective phonon-spin qubit coupling in the diamond cavity thus reaches up to $g_{\rm pe}/(2\pi)\approx 0.1 g_{\rm orb}/(2\pi)=2.4$\,\si{\mega\hertz}. This diamond cavity furthermore offers the possibility to increase the efficiency of optical addressing of the diamond AAs by concentrating light of a vacuum wavelength $\lambda_{\rm opt}=732$\,\si{\nano\meter} into an optical mode that is spatially overlapping with the cavity mechanical mode. The high calculated optical quality factor $Q_{\rm opt}=10^6$ can be used to increase the efficiency of optical addressing of the diamond AA as discussed further in Sec.\,\ref{sec:VI}. 

In summary, we have designed (opto)mechanical cavities that sustain mechanical modes whose zero-point strain fluctuations enable \textit{strong coupling} between an AA spin and a single quantum of mechanical motion. The feasibility of such devices marks an important practical step towards transducers relying on spin-strain interactions. Having established the achievable values of couplings and decay rates governing the dynamics of the system, we now proceed to analyse the numerical results of our quantum-sate transduction protocol.

%\begin{table}[h]
%\caption{Parameters used for the calculations of the transduction of a quantum state from the SC qubit to the nuclear spin (unless specified otherwise). }
%   \centering
%    \begin{tabular}{c|c}
%    Quantity & Magnitude\\
%    \hline
%        $g_{\rm scp}$ & $2\pi\times$ 50 MHz\\
%        $g_{\rm pe}$ & $2\pi\times$ 1 MHz\\
%        $\gamma_{\rm sc}$ & $2\pi\times$ 10 kHz \\
%        $\gamma_{\rm p}$ & $2\pi\times$ 100 Hz\\
%        $\gamma_{\rm e}$ & $2\pi\times$ 10 kHz\\
%        $\gamma_{\rm n}$ & $2\pi\times$ 1 Hz.
%    \end{tabular}
%   \label{tab:parameters}
%\end{table}
%((( dirk comment: why assume such a small SC frequency? normally ppl use like 5-10 GHz.. and your SC Q value is quite low; i believe ~ 1M is doable in planar geometries (see eg LL), or are you worried about the phonon mode to have lower Q at high frequency? I would hope to eliminate it altogether, i.e., if the expectation of the phonon field is in the grnd state)))

\section{Numerical Analysis of SC-Emitter Quantum State Transfer}\label{sec:V}
\begin{figure}
    \centering
    \includegraphics[scale=1]{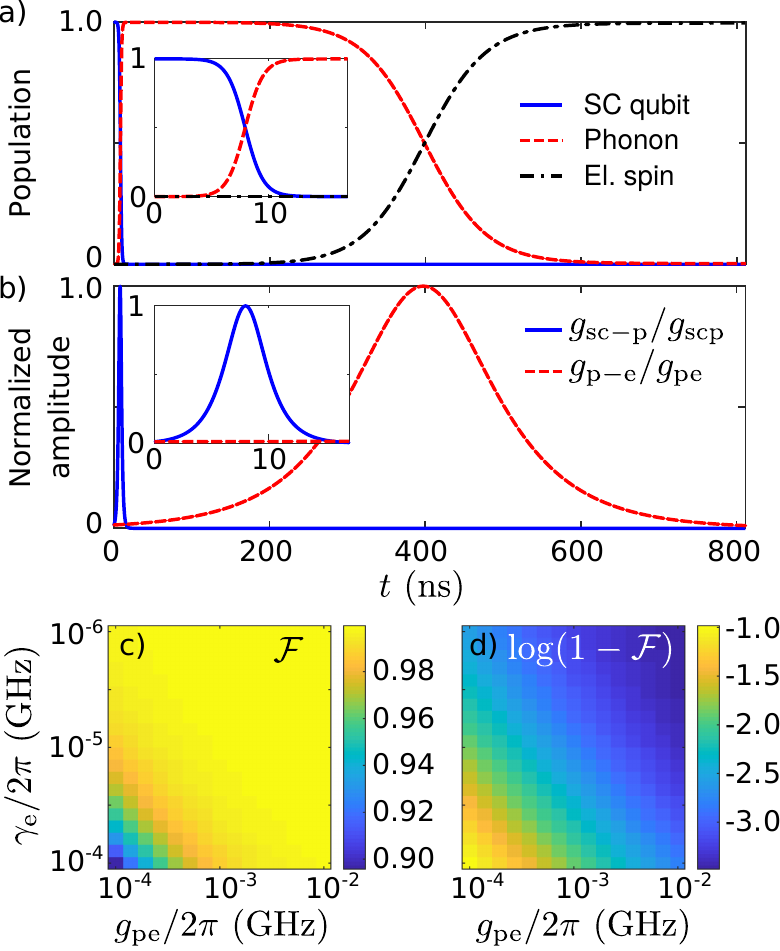}
    \caption{State transfer from the SC qubit to the electron spin. The system begins in the excited state of the SC qubit, then evolves according to the master equation Eq.\,\eqref{eq:mastereqdirect}. The time-dependent populations of the SC qubit (full blue), the cavity phonon (red dashed), and the electron spin qubit (black dash-dotted) are plotted as a function of time in (a). After initializing the system, we apply the series of pulses shown in (b), as described in the text, and let the system evolve until the state is transferred to the electron spin. (c) The state-transfer fidelity $\mathcal{F}$, and (d) ${\log}(1-\mathcal{F})$. Both $\mathcal{F}$, and ${\log}(1-\mathcal{F})$ are calculated as a function of the phonon-electron-spin coupling $g_{\rm pe}$ and the electron-spin dephasing rate $\gamma_{\rm e}$. We maximize $\mathcal{F}$ for each pair of parameters [$\gamma_{\rm e}$, $g_{\rm pe}$] by adjusting the delays between the respective pulses applied to drive the system dynamics. We consider $\gamma_{\rm p}/(2\pi)=10^{-7}$\,GHz (corresponding to the mechanical quality factor $Q\sim 10^{7}$) in the calculations.}
    %\cmmb{[T: I would show only the result for the "direct scheme"]}
    \label{fig:rabiscsp}
\end{figure}

%- outline goal 
%- how to achieve (tunable couplings in time, minimize time ... ) 
%- any experimental considerations/what is the limiting factor? 
%- what is the solution for tunable couplings (show fig?)
%- describe results etc ... 
%- talk about parameter sweep 
As discussed in Sec.\,\ref{sec:IV}, mechanical resonators of high quality factors exceeding $Q\sim 10^7$ have been demonstrated experimentally~\cite{maccabe2019phononic}. The limiting time-scale for high-fidelity state transfer is therefore the decoherence of the SC and electron spin qubits, so it is necessary to transfer the SC population rapidly into the phononic mode. The long-lived phonon then allows transduction into the AA electron spin levels of the emitter, where the qubit can be addressed optically, or is further transferred to the quantum memory - the nuclear spin \cite{bradley2019register, nguyen2019nuclearoptics}. We numerically evaluate the master equation, and show the results of the time evolution of such a system in Fig.\,\ref{fig:rabiscsp} (a). In our simulation we consider $g_{\rm scp}/(2\pi)=50$\,\si{\mega\hertz}, $g_{\rm pe}/(2\pi)=1$\,\si{\mega\hertz}, $\gamma_{\rm sc}/(2\pi)=10^{-5}$\,\si{\giga\hertz} \cite{Kjaergaard_2019}, $\gamma_{\rm p}/(2\pi)=10^{-7}$\,\si{\giga\hertz}, $\gamma_{\rm e}/(2\pi)=10^{-5}$\,\si{\giga\hertz}. The SC qubit is initialized in the excited state while the rest of the system is considered to be in the ground state. We let the system evolve in time and apply the series of control pulses [Eq.\,\eqref{eq:pulsescp} and \eqref{eq:pulsepe} shown in Fig.\,\ref{fig:rabiscsp}(b) as a blue line and a red dashed line, respectively] to transfer the initial population of the SC qubit (full blue line) sequentially to the phonon (red dashed line), and the electron spin (black dash-dotted line), as shown in Fig.\,\ref{fig:rabiscsp}(a). 

To further analyse the transduction we calculate the state-transfer fidelity $\mathcal{F}$ defined in Eq.\,\eqref{eq:fidelity} as a function of the phonon-spin coupling $g_{\rm pe}$ and the electron-spin dephasing rate $\gamma_{\rm e}$. We vary $g_{\rm pe}/(2\pi)$ in the range from $100$\,kHz, representing a conservative estimate of the phonon-spin coupling rate, to $10$\,MHz which exceeds the value we estimate for the silicon phononic cavity by an order of magnitude. 
The role of the electron-spin coherence on the overall state-transfer fidelity [together with the infidelity $\log{(1-\mathcal{F})}$] is shown in Fig.\,\ref{fig:rabiscsp}(c)  [Fig.\,\ref{fig:rabiscsp}(d)]. When calculating $\mathcal{F}(g_{\rm pe}, \gamma_{\rm e})$ we set $\gamma_{\rm p}/(2\pi)=10^{-7}$\,\si{\giga\hertz}, i.e. we consider a high-quality resonance of the phononic cavity. We consider $\gamma_{\rm e}/(2\pi)=10^{-4}$\,GHz as a conservative upper bound of the electron-spin decoherence rate. However, progress in quantum technology indicates that the lower value considered in our calculations, $\gamma_{\rm e}/(2\pi)=10^{-6}$\,\,\si{\giga\hertz}, can be achieved in state-of-the-art systems \cite{nguyen2019strainsi}. Our calculation shows that for high transfer fidelity (infidelity of less than $\sim 1$\%) the electron-spin decoherence rate should not exceed $\gamma_{\rm e}/(2\pi)\approx 10^{-5}$\,\si{\giga\hertz}, well within the experimentally accessible range, indicating electro-mechanical state transfer is potentially achievable in present systems.

\section{Quantum Interfacing}\label{sec:VI}
The AA electron-spin qubit serves as the network bus, mediating coupling to not just phonons as well as photons and nuclear spins. In particular, the spin-dependent optical transitions enable photon-mediated coupling of the quantum device to, for example, distant quantum memories in a quantum network, as illustrated in Fig.\,\ref{fig:schematicfig1}(a).  
One approach would be to perform spin-to-photon conversion by optically addressing the electron spin after the SC qubit has been transduced to it. This could be performed via a variety of spin-photon interfacing procedures, including direct optical excitation of the quantum emitter \cite{becker2018alloptical}, or a spin-photon controlled-phase gate mediated by a cavity mode. However, the fidelity of this approach is intrinsically limited by the achievable emitter-cavity cooperativity and the detuning between spin states. In particular, current experiments have demonstrated entanglement fidelities of 0.94 and heralding efficiencies of 0.45 \cite{bhaskar2019experimental}.
The photon loss associated with this direct spin-to-photon transduction also destroys the quantum state that was to be transported.

An alternative approach first entangles a nearby nuclear spin with the quantum network target. This can be achieved using the procedure of repeat-until-succeed optical heralding developed for deterministic state teleportation \cite{Pfaff532, humphreys2018deterministic, rozpkedek2019near, bhaskar2019experimental}. This scheme never actually transduces the SC qubit to the optical domain and thus avoids photon transmission losses. Instead, electron-nuclear spin gates can be used to teleport the qubit across a quantum network. This second approach can achieve near-unity state-transfer fidelity and efficiency provided that entangled qubit pairs shared between nodes of the quantum network can be prepared on demand. This preparation of on-demand entanglement has been recently realized for diamond NV centers~\cite{humphreys2018deterministic}. 
%( Science review by Wehner and Hanson et al from 2019 for example;  Bhaskar  - Englund - Loncar - Lukin 2019 ArXiv). 
So far, spin-spin teleporation fidelities of up to 0.84 have been reported~\cite{Hensen2015, humphreys2018deterministic} using the NV center in diamond. Ongoing experimental and theoretical advances promise to enable near-unity teleportation fidelity, including through environmentally-insensitive quantum emitters (such as the SiV considered in this work) and entanglement schemes to improve noise-and error-resilience.
%%TOMAS END

The hyperfine interactions of the electron spin with nearby spins of nuclear isotopes is often an unwanted source of electron-spin decoherence hindering the ability to maintain and control the electron-spin qubits over long time scales. Dynamical decoupling techniques \cite{de2010universal, farfurnik2015decoupling} have been applied to mitigate this decoherence and reach $\sim1$\,\si{\milli\second} to $\sim10$\,\si{\milli\second} coherence times in SiV systems \cite{christle2015isolated, sukachev2017sivspinqubit, becker2018alloptical}. However, recent theoretical and experimental work shows that the nuclear spins can be used as a resource as their quantum state can be selectively addressed and controlled via the quantum state of the electron spin itself \cite{taminiau2014universal,Waldherr2014, bradley2019register} with high fidelity. Combined with the extraordinarily long (exceeding $\sim$1\,\si{\second}) coherence times of these nuclear spins, it has been proposed that the nuclear-spin bath could serve as a quantum register \cite{taminiau2014universal, bradley2019register, nguyen2019nuclearoptics} and could store quantum states and thus serve as a QM. In Appendix\,\ref{app:nucspin} we describe how the protocol developed in \cite{bradley2019register} can implement a quantum SWAP gate allowing for state transfer from the electron-spin qubit to a single nuclear spin of a nearby $^{13}$C atom. Assuming electron-spin pure dephasing of $\gamma_{\rm e}/(2\pi)=10$\,\si{\kilo\hertz}, nuclear-spin pure dephasing of $\gamma_{\rm n}/(2\pi)=1$\,\si{\hertz}, a moderate electron-spin hyperfine coupling $A_{\parallel}=500$\,\si{\kilo\hertz}, and a conservative value of an external microwave drive Rabi frequency $\Omega_{\rm mw}/(2\pi)\approx 3.9$\,\si{\kilo\hertz}, we estimate that the state-transfer fidelity $\mathcal{F}_{\rm en}$ of this process could reach $\mathcal{F}_{\rm en}\approx 0.9975$. 

% scalability paragraph matt 
The compactness of this diamond QM further opens up the possibility to scale the system. Using a mechanical or microwave switching network, each SC qubit could be selectively coupled to a large number of mechanical cavities depending on the experimental architecture. As each additional coupled cavity introduces a decay channel, low-loss high-isolation switching is required. As an example, we consider a pitch-and-catch scheme could [Appendix\,\ref{app:indirect}] wherein the quantum state is launched into a mechanical waveguide with controllable coupling to many phononic resonators. For high fidelity state transfer $\mathcal{F} > 0.99$, the total insertion loss of all switches must remain below 0.04 dB, which may prove experimentally challenging. 
%[[To tomas: is there a rate picture for this?]  
Considering experimentally achievable AA densities, we estimate that about 10 AAs could be individually addressed within the mechanical mode volume of $\sim 10^{7}$\,\si{\nano\meter^3} of each waveguide. These can be individually optically addressed due to their inhomogeneous optical and microwave transition distribution \cite{bersin2019individual, neuman2019selective}, induced by natural variations in local static strain within the diamond crystal. 

Each color center enables high-fidelity coupling to $\sim 10$ nuclear spins \cite{bradley2019register}. Allowing for, say, 10 parallel QM interconnects from the QPU would thus provide a total QM capacity of about $\sim 10\times 10\times 10 =$ kqubits. Introducing spatial multiplexing (e.g., using microwave switches) would multiply the QM capacity further. 

As we show in Appendix\,\ref{app:molmersor}, the proposed architecture coupling a large number of electron-spin qubits to a shared mechanical mode further opens the opportunity for efficient phonon-mediated spin-entangling quantum gates \cite{molmer1999multiparticleentanglement, sorensen1999quantumcomputation1, kuzyk2018scaling, li2019honeycomb}. These gates enable preparation of highly entangled many-spin states, such as the Greenberger–Horne–Zeilinger (GHZ) state, that can serve as resources for further quantum-state manipulation. Specifically, an $N$-electron-spin GHZ state coupled to the same phononic cavity could would increase the phase sensitivity to strain $N$-fold \cite{Leibfried1476}. Thus, a GHZ state prepared in advance of the SC-to-spin transduction would speed up the controlled phase gates, $N$ times speedup for a $N-$spin GHZ state. Combined with local gates acting on the spins and the phonon, a GHZ state could be used to boost the speed and fidelity of the SWAP gate.

\section{Conclusions and Outlook}\label{sec:VII}
%[Dirk t mention : enhancing rates through GHZ-enhanced and/or other color centers / stronger magnetic field;  open questions and problems that need more investigation, including experimental evidence of coherent MW spin-photon coupling; measurements of diamond-phonon cavities;  multiplexing approaches; extension for QRAM; ]

%We have introduced an architecture for high-bandwidth, high-fidelity quantum state transduction between superconducting microwave AA spins  qubits at rates far exceeding intrinsic system decay and decoherence. The resulting hybrid architecture combines the favorable attributes of quantum memories with SC quantum information processors, enabling a wide range of functionalities currently unavailable to a stand-alone superconducting or spin-based architectures. The long coherence times of the spins allows them to operate as a memory, storing quantum information for orders of magnitude longer than the best superconducting devices. Many mechanical cavities can be interfaced with the MW circuits and the proposed system is thus highly scalable extensible to a a high-capacity quantum memory. 

We introduced an architecture for high-bandwidth, high-fidelity quantum state transduction between superconducting microwave AA spin qubits at rates far exceeding intrinsic system decay and decoherence. The resulting hybrid architecture combines the favorable attributes of quantum memories with SC quantum information processors, enabling a wide range of functionalities currently unavailable to a stand-alone superconducting or spin-based architectures. 

Strong coupling of a single defect center spin to a high-quality mechanical cavity, the key element of our proposal, remains to be experimentally demonstrated. Nevertheless, our analysis shows experimental feasibility of the proposal in state-of-the-art mechanical systems. Further experimental challenges exist in the demonstrations of controllable electro-mechanical and mechanical-mechanical couplers that are necessary for the cascaded state transfer. Rapid development of micro-mechanical systems indicates that the above mentioned experimental challenges can be solved in the foreseeable future.  

Looking further ahead, reaching fault tolerant quantum information processing will likely require gate and measurement errors below 0.1\%. Fortunately, there appear to be several avenues to speed up, and thus improve, the fidelity of quantum state transfer between phonon to spin encoding. These include further strain concentration (e.g. through thinner diamond patterning), identifying different AAs with increased strain coupling, state distillation, and the use of pre-prepared spin GHZ states---and to that end, fast and reliable spin-entangling protocols must be developed both theoretically and experimentally.
 
To summarize our key results, the SC-AA hybrid architecture combines the complementary strengths of SC circuit quantum computing and artificial atoms, realizing the essential elements of an extensible quantum information processing architecture.%: see Table~\ref{tab:summary}. 
There are, of course, components that need to be realized and assembled into one system, which will diminish certain performance metrics, at least near-term. Nonetheless, even our basic performance considerations show that these different capabilities -- QPU, QM, bus, and quantum network port -- should leverage distinct physical modalities in a hybrid system, much like a classical computing system.

\section*{Acknowledgements}
This material is based upon work supported by the U.S. Department of Energy, Office of Science, Basic Energy Sciences (BES), Materials Sciences and Engineering Division under FWP ERKCK47 (T.N., M.T., P.N. and D.E.)`Understanding and Controlling Entangled and Correlated Quantum States in Confined Solid-state Systems Created via Atomic Scale Manipulation'. The authors (T.N., M.T., P.N.) acknowledge the `Photonics at Thermodynamic Limits' Energy Frontier Research Center funded by the U.S. Department of Energy, Office of Science, Office of Basic Energy Sciences under Award Number DE-SC0019140 that supported computational approaches used here. D.E. acknowledges partial support from a Bose Research Fellowship. P. N. is a Moore Inventor Fellow supported by the Gordon and Betty Moore Foundation. This work is based upon work supported by the U.S. Department of Energy, Office of Science, Advanced Scientific Computing Research (ASCR) under FWP 19-022266 (M.E. and L.H.) `Quantum Transduction and Buffering Between Microwave Quantum Information Systems and Flying Optical Photons in Fibers'. M.E. performed this work, in part, at the Center for Integrated Nanotechnologies, an Office of Science User Facility operated for the U.S. Department of Energy (DOE) Office of Science. M.E. and L.H. were supported by the Laboratory Directed Research and Development program at Sandia National Laboratories, a multimission laboratory managed and operated by National Technology and Engineering Solutions of Sandia, LLC., a wholly owned subsidiary of Honeywell International, Inc., for the U.S. Department of Energy's National Nuclear Security Administration under Contract No. DE-NA-003525. This paper describes objective technical results and analysis. Any subjective views or opinions that might be expressed in the paper do not necessarily represent the views of the U.S. Department of Energy or the United States Government.

\bigskip
T.N., M.E., and M.T. contributed equally to this work.

\appendix

\section{Transduction from a SC qubit to a spin qubit via a waveguide}\label{app:indirect}

The main text discussed a scheme where the SC qubit is directly electro-mechanically coupled to a phononic cavity. Alternatively, the mechanical mode of the phononic cavity can be coupled to the microwave circuit via a microwave or phononic waveguide. For example, this waveguide may serve as an interconnect between a SC qubit of a quantum computer which is physically separated from the phononic cavity across large distance, or it might represent a guided phonon wave connecting a piezoelectric coupler (interdigital coupler - IDT) with a discrete high-$Q$ mechanical mode of a phononic cavity surrounded by a phononic crystal.   

We break down the transduction of the qubit stored in the SC device to the spin and describe in this appendix the `pitch-and-catch' state transfer of the SC state to the mechanical resonator via the waveguide. The transduction of the quantum state stored in the phonon into the electron spin via an effective Jaynes-Cummings interaction can be performed as described in the main text.  

\subsection{State transfer from the SC qubit to the phononic cavity via a waveguide}

\begin{figure}
    \centering
    \includegraphics[scale=0.95]{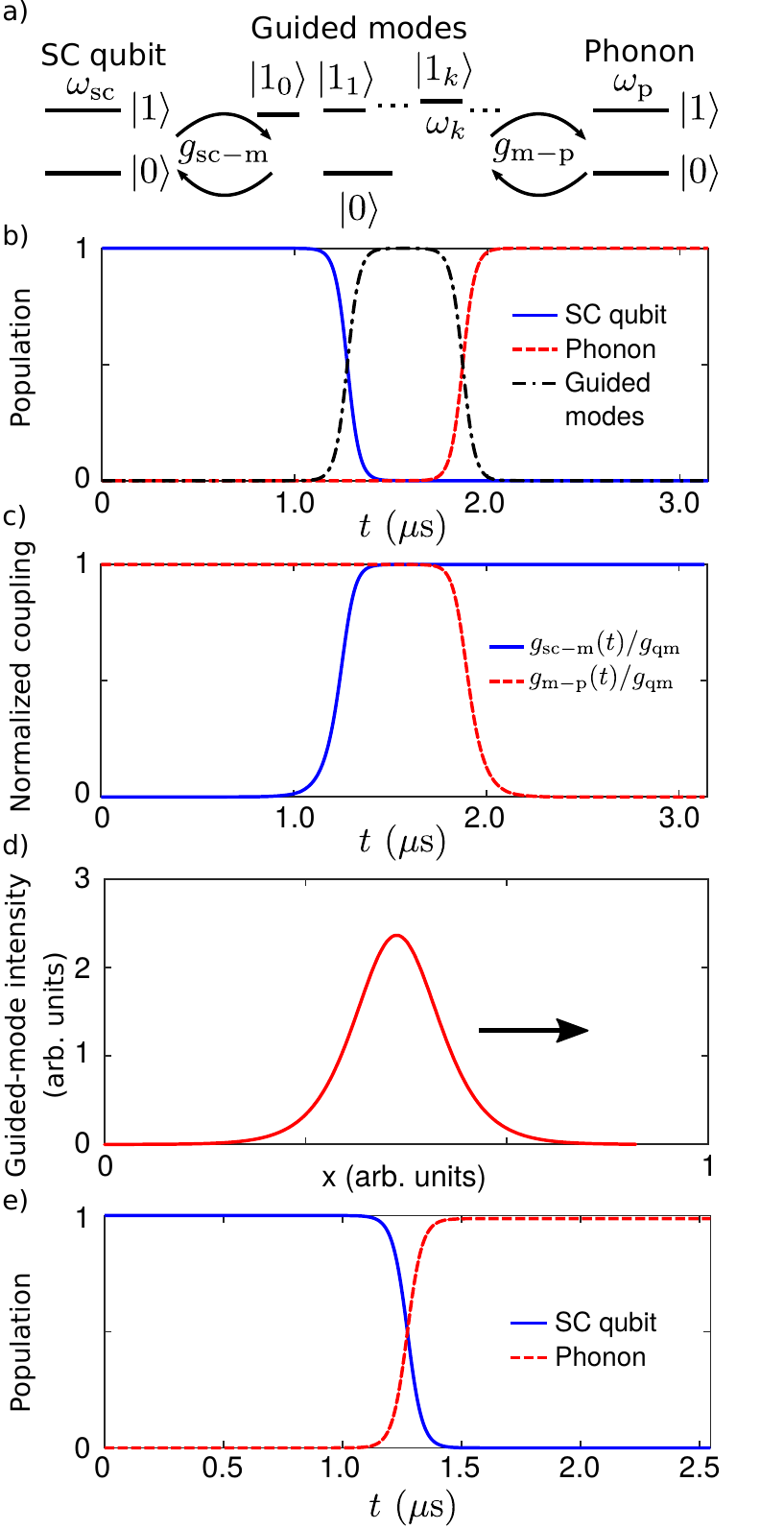}
    \caption{Pitch-and-catch scheme for state transfer between the SC qubit and the phononic cavity. (a) The general scheme describing a SC qubit (of frequency $\omega_{\rm sc}$) coupled to a continuum of waveguide modes $k$ of their respective frequencies $\omega_k$ via time-dependent coupling rate $g_{\rm sc-m}$ that mediates the state transfer to the phononic cavity of frequency $\omega_{\rm p}$ via a time-dependent coupling rate $g_{\rm m-p}$. (b) Populations of the SC qubit (full blue), the phonon (red dashed), and the propagating photon (black dash-dotted) as a function of time. (c) The time-dependent coupling rates $g_{\rm sc-m}$ and $g_{\rm m-p}$ applied to control the state transfer. (d) The SC qubit releases a symmetrical photon wave packet (photon intensity shown as a function of distance along the waveguide) propagating towards the receiving phononic cavity (shown by the arrow). (e) The state transfer as in (b) but described using the master-equation approach [Eq.\,\eqref{eq:mastercascade}]. Only the SC-qubit population (blue line) and the phonon population (dashed red line) are shown as the photons are eliminated from the dynamics. We disregard the time delay caused by the photon propagation along the finite-length waveguide for simplicity.}
    \label{fig:releasecatch}
\end{figure}

As schematically shown in Fig.\,\ref{fig:releasecatch}(a), we assume that the SC qubit is coupled to a waveguide which is electro-mechanically coupled to the phononic cavity (or, alternatively, to a phononic waveguide mechanically coupled to a discrete mechanical mode of a cavity). Such a system can be described by the following Hamiltonian \cite{Milonni1983recurrences}:
\begin{align}
    H_{\rm sc-m-p}&=\hbar\omega_{\rm sc}\sigma^\dagger_{\rm sc}\sigma_{\rm sc}+\hbar\omega_{\rm p}b^\dagger b +\sum_k \hbar \omega_{k}a^\dagger_k a_k  \nonumber\\
    &+\sum_{k}\hbar g_{\rm sc-m}(t) (\sigma^\dagger_{\rm sc} a_k+\sigma_{\rm sc}a_k^\dagger)\nonumber\\
    &+\sum_k \hbar g_{\rm m-p}(t) (b^\dagger a_k+b a^\dagger_k). 
\end{align}
where $a_k$ ($a^\dagger_k$) are the annihilation (creation) operators of a waveguide mode $k$ of frequency $\omega_k$. The SC qubit and the mechanical mode are coupled to the waveguide via a controllable time-dependent coupling $g_{\rm sc-m}(t)$ \cite{chen2014tunecoupler, Bienfait2019transferSAW, geller2015tunable} and $g_{\rm m-p}(t)$, respectively. The coupling $g_{\rm m-p}$ can be either realized as a tunable IDT coupler, or as a tunable mechanical interconnect that could be  e.g. based on interferometric modulation of coupling in analogy with optical implementations \cite{Tanaka2007,Kumar2011}, although an implementation of such a controllable phononic coupler is yet to be demonstrated.  

The quantum state stored in the SC device can be released into the waveguide and subsequently absorbed by the phononic cavity. To accomplish the pitch-and-catch state transfer with high fidelity we need to ensure that the processes of phonon emission by the SC qubit and phonon absorption by the phononic cavity are mutually time-reversed. To that end the pulse emitted by the SC qubit has to be time-symmetrical and the couplings have to fulfill $g_{\rm sc-m}(t)=g_{\rm m-p}(-[t-\tau])$ \cite{cirac1997transfer}, where $\tau$ is the delay time due to the finite length of the waveguide. 

For concreteness, we consider a waveguide of length $L$ supporting phononic (or electromagnetic) modes of the form $\propto \cos(k_j x)$, with $k_j=(N_0+j)\pi/L$ and $x$ being a position along the waveguide, where $N_0$ is a mode number that in connection with the waveguide length $L$ and the mode velocity $c$ (assuming a linear dispersion) determines a central frequency of the selected set of modes. This function may represent a vector potential in a MW transmission line or e.g. a mechanical displacement of a phononic wave. The free spectral range of this finite waveguide is $\delta=c\pi/L$ and the spontaneous decay rate of each qubit into the waveguide modes occurs with the rate (assuming time-independent $g_{\rm scm}=g_{\rm mp}\equiv g_{\rm qm}$):
\begin{align}
    \kappa_{\rm sc}=\frac{2\pi g_{\rm q-m}^2}{\delta}.\label{eq:defkapsc}
\end{align}
The objective of releasing a perfectly symmetrical microwave pulse in a form proportional to ${\rm sech}(\kappa_{\rm sc}t/2)$ can be achieved if we modulate the coupling constant in time via an electromechanical coupler \cite{Bienfait2019transferSAW}:
\begin{align}
    g_{\rm sc-m}(t)=g_{\rm qm}\sqrt{\frac{e^{\kappa_{\rm sc}t}}{1+e^{\kappa_{\rm sc}t}}},
\end{align}
The wave packet released by the superconducting qubit can then be fully absorbed by the phononic cavity if the time-reversed delayed coupling is:
\begin{align}
    g_{\rm m-p}(t)=g_{\rm qm}\sqrt{\frac{e^{-\kappa_{\rm sc}(t-\tau)}}{1+e^{-\kappa_{\rm sc}(t-\tau)}}}.\label{eq:defgmp}
\end{align}
We demonstrate the pitch-and-catch scheme in Fig.\,\ref{fig:releasecatch}(b-d). Figure\,\ref{fig:releasecatch}(b) shows the time-dependence of the populations of the SC qubit (full blue), the phonon (red dashed), and the MW photon in the waveguide (black dash-dotted). As shown, the pitch-and-catch scheme leads to an almost perfect transfer of population from the SC qubit to the phonon [final phonon population $\langle \sigma^\dagger_{\rm p}(t_{\rm end})\sigma_{\rm p}(t_{\rm end})\rangle\approx 1$]. The sequence of time-dependent couplings shown in Fig.\,\ref{fig:releasecatch}(c) first releases a fully symmetrical propagating wave packet [a snapshot of the photon intensity is shown in Fig.\,\ref{fig:releasecatch}(d) as a function of position along the waveguide] and is subsequently perfectly absorbed by the receiving qubit. We cast the model outlined above into the form of a master equation \cite{gardiner1993driving} for the density matrix $\rho_{\rm scp}$ describing the SC qubit and the phononic cavity in the single-excitation basis, but only effectively accounting for the modes of the MW waveguide:
\begin{align}
    \frac{\partial \rho_{\rm scp}}{\partial t}&=-\frac{{\rm i}}{\hbar}[H_{\rm sc}+H_{\rm p}, \rho_{\rm scp}]\nonumber\\
    +&\kappa_{\rm sc}(t)\mathcal{L}_{\sigma_{\rm sc}}(\rho_{\rm scp})+\gamma_{\rm sc}\mathcal{L}_{\sigma_{\rm sc}}(\rho_{\rm scp})\nonumber\\
    +&\kappa_{\rm p}(t)\mathcal{L}_{b}(\rho_{\rm scp})+\gamma_{\rm p}\mathcal{L}_{b}(\rho_{\rm scp})\nonumber\\
    +&\sqrt{\kappa_{\rm p}(t)\kappa_{\rm sc}(t)}\nonumber\\
    \times&\left(  e^{{\rm i}\phi}[\sigma_{\rm sc}\rho_{\rm scp}, b^\dagger]+e^{-{\rm i}\phi}[b, \rho_{\rm scp}\sigma^\dagger_{\rm sc}]  \right),\label{eq:mastercascade}
\end{align}
with $H_{\rm sc}=\hbar\omega_{\rm sc}\sigma^\dagger_{\rm sc}\sigma_{\rm sc}$, and $H_{\rm p}=\hbar\omega_{\rm p}b^\dagger b$. It is understood that the density matrix of the phonon is evaluated at a later time $t+\tau$ (in the following we always set $\tau=0$\,s for simplicity) and the phase accumulated due to the propagation of the photon wave packet is absorbed in the definition of $\phi$. The respective time-dependent decay rates are given by:
\begin{align}
    \kappa_{\rm sc}&=\frac{2\pi g_{\rm sc-m}^2(t-\tau_{\rm pc})}{\delta},\\
    \kappa_{\rm p}&=\frac{2\pi g_{\rm m-p}^2(t-\tau_{\rm pc})}{\delta},
\end{align}
in accordance with Eqs.\,\eqref{eq:defkapsc}-\eqref{eq:defgmp}. We consider that the pulses are applied at a later time $\tau_{\rm pc}$ to ensure smooth dynamics. The resulting time-dependent populations shown in Fig.\,\ref{fig:releasecatch}(e) perfectly capture the pitch-and-catch scheme described previously in the framework of Schr\"{o}dinger equation [cf. populations in Fig.\,\ref{fig:releasecatch}(b)]. 
Last we note that by effectively eliminating the waveguide we neglect the waveguide propagation losses that could further decrease the state-transfer fidelity. Neverheless, we estimate that for phonon decay rates $\sim 1$\,\si{\hertz} achieved in state-of-the-art acoustical systems, speed of sound $c\sim 10^3$\,\si{\meter\second^{-1}}, and waveguide length $L\sim 1$\,\si{\milli\meter}, the propagation losses are so small to result in near-unity transmission $\sim e^{-10^{-6}}\approx 0.999999$. 
We integrate the master-equation description of the pitch-and-catch scheme [Eq.\,\eqref{eq:mastercascade}] to describe the full dynamics of the state transfer from the SC qubit to the electron spin and show the result of the state-transfer protocol in Fig.\,\ref{fig:releasecatch}(e).

\section{Effects of strain on SiV negative center}\label{app:strain}

\begin{figure}
    \centering
    \includegraphics[scale=1]{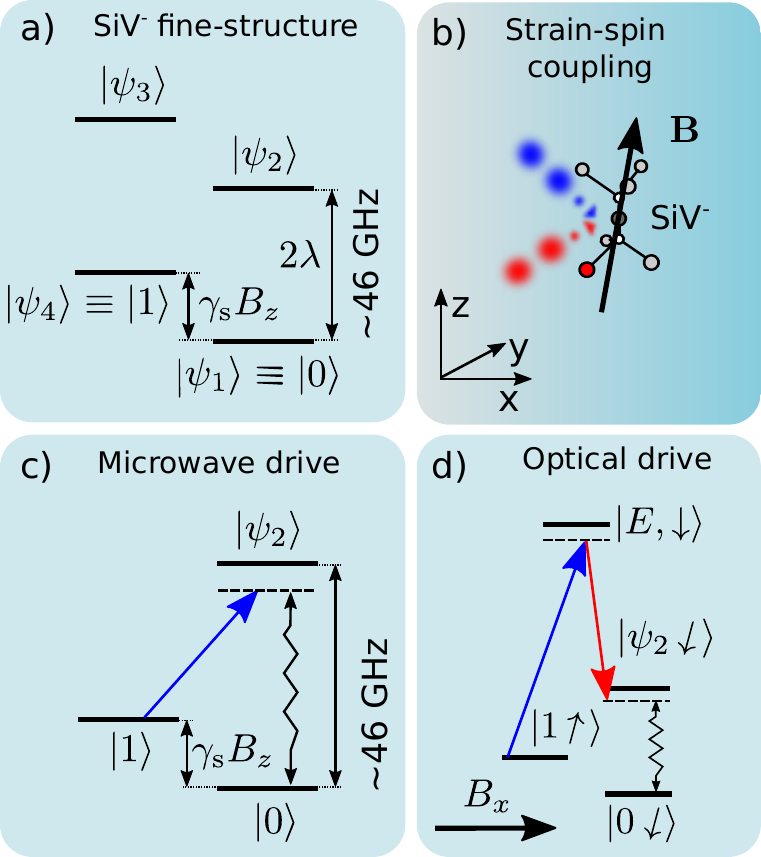}
    \caption{Coupling between fine-structure states of a SiV$^-$ defect and strain. (a) Fine structure states described in Appendix\,\ref{app:strain}. (b) External magnetic field ${\bf B}$ and optical drive is applied to the defect to enable the strain-spin coupling. (c) A microwave-based Raman scheme: states $|1\rangle$ and $|\psi_2\rangle$ can couple via magnetic field whereas states $|0\rangle$ and $|\psi_{2}\rangle$ are coupled via strain. (d) An optical Raman scheme exploiting a virtual excitation via an optically-active excited state can be used to induce an effective strain-mediated coupling between $|0\rangle$ and $|1\rangle$.}
    \label{fig:strainspin}
\end{figure}

The effects of strain on a SiV$^-$ center have been considered in the literature \cite{meesala2018strainsiv, nguyen2019strainsi} theoretically and experimentally. The theory predicts that the strain effects can be divided into three categories according to the transformation properties of the strain field under symmetry operation of the D$_{3{\rm d}}$ symmetry group. based on symmetry, the strain can be classified as $\epsilon_{\rm A_{1g}}$, $\epsilon_{\rm E_{gx}}$ and $\epsilon_{\rm E_{gy}}$. These strain components then give rise to the longitudinal, $\alpha$, and transverse, $\beta$ and $\gamma$, strain coupling to the spin-orbit states of the color center:
\begin{align}
    \alpha&=t_\perp (\epsilon_{xx}+\epsilon_{yy})+t_\parallel \epsilon_{zz}\sim\epsilon_{\rm A_{1g}},\\
    \beta&=d(\epsilon_{xx}-\epsilon_{yy})  + f\epsilon_{zx}\sim\epsilon_{\rm E_{gx}},\label{eq:strainbeta}\\
    \gamma&=-2 d(\epsilon_{xy})  + f\epsilon_{yz}\sim\epsilon_{\rm E_{gy}},\label{eq:straingamma}
\end{align}
where $z$ is oriented along the high-symmetry axis of the defect $[111]$,  $x$ is oriented along $[\bar{1}\bar{1}2]$, and $y$ is defined by $[\bar{1}10]$. The respective values of the constants $t_\parallel$, $t_\perp$, $d$, and $f$ have been estimated to be in the range of $1$\,PHz to $2$\,PHz (we transform the relevant tensor components into the coordinate system defined by $[100]$, $[010]$, and $[001]$ in Appendix \,\ref{app:transformed}). We will use these values to estimate all necessary constants to design a potential transducer. As has been shown \cite{meesala2018strainsiv, maity2018alignment, lemonde2018phononnetworks, nguyen2019strainsi, Maity2020straincontrol}, the A$_{\rm 1g}$ strain uniformly shifts all the fine-structure-state energies and we thus disregard its effects in the following discussion. 

We further consider that the Hamiltonian of the fine-structure states of a SiV$^-$ in a longitudinal magnetic field is (neglecting the Jahn-Teller effect and the Orbital Zeeman effect \cite{lemonde2018phononnetworks}, for simplicity):
%\begin{widetext}
\begin{align}
H_{\rm tot}=\left(
\begin{array}{cccc}
 B_z \gamma _{\rm S} & 0 & -{\rm i} \lambda  & 0 \\
 0 & -B_z \gamma _{\rm S} & 0 & {\rm i} \lambda  \\
 {\rm i} \lambda  & 0 & B_z \gamma _{\rm S} & 0 \\
 0 & -{\rm i} \lambda  & 0 & -B_z \gamma _{\rm S} \\
\end{array}
\right).\label{eq:hamfine}
\end{align}
%\end{widetext}
The Hamiltonian is expressed in the basis of spin-orbit states $\{|e_y\uparrow \rangle, |e_y\downarrow \rangle, |e_x\uparrow \rangle, |e_x\downarrow\rangle \}$ \cite{meesala2018strainsiv, nguyen2019strainsi}. Here $\lambda$ is the spin-orbit coupling strength ($\lambda/(2\pi \hbar)\approx 23$\,GHz), $B_z$ is the magnetic field applied along the high-symmetry axis of the defect, and $\gamma_{\rm S}/(2\pi)\approx  28$\,GHz/T is the spin gyromagnetic ratio. The Hamiltonian in Eq.\,\eqref{eq:hamfine} can be diagonalized to obtain the eigenfrequencies:
\begin{align}
   \nu_1&=-B_z \gamma _{\rm S}-\lambda ,\\
   \nu_2&=-B_z \gamma _{\rm S}+\lambda ,\\
   \nu_3&=B_z \gamma _{\rm S}+\lambda ,\\
   \nu_4&=B_z \gamma _{\rm S}-\lambda.
\end{align}
and the corresponding eigenstates:
\begin{align}
    |\psi_1\rangle&=\frac{1}{\sqrt{2}}(-{\rm i} |e_y\downarrow \rangle+|e_x\downarrow\rangle),\\
    |\psi_2\rangle&=\frac{1}{\sqrt{2}}({\rm i} |e_y\downarrow \rangle+|e_x\downarrow\rangle),\\
    |\psi_3\rangle&=\frac{1}{\sqrt{2}}(-{\rm i} |e_y\uparrow \rangle+|e_x\uparrow\rangle),\\
    |\psi_4\rangle&=\frac{1}{\sqrt{2}}({\rm i} |e_y\uparrow \rangle+|e_x\uparrow\rangle).
\end{align}
The structure of the fine-structure states is schematically depicted in Fig.\,\ref{fig:strainspin}(a). 
The two lowest-energy states $|\psi_4\rangle$ and $|\psi_1\rangle$ can be conveniently used as the spin-qubit states.   
In this basis the transverse-strain Hamiltonian becomes:
\begin{align}
H_{ \mathcal{\beta}}=\left(
\begin{array}{cccc}
 0 & {\beta} & 0 & 0 \\
 {\beta} & 0 & 0 & 0 \\
 0 & 0 & 0 & {\beta} \\
 0 & 0 & {\beta} & 0 \\
\end{array}
\right)
\end{align}
and 
\begin{align}
H_{ \gamma}=\left(
\begin{array}{cccc}
 0 & {\rm i} \gamma & 0 & 0 \\
 -{\rm i} \gamma & 0 & 0 & 0 \\
 0 & 0 & 0 & {\rm i} \gamma \\
 0 & 0 & -{\rm i} \gamma & 0 \\
\end{array}
\right),
\end{align}
where $\beta$ and $\gamma$ are the strain components shown in Eq.\,\eqref{eq:strainbeta} and Eq.\,\eqref{eq:straingamma}. The spin degree of freedom thus cannot be flipped by the sole application of a transverse strain. Considering that the transition $|\psi_1\rangle\leftrightarrow|\psi_4\rangle$ is spin-forbidden and the states $|\psi_1\rangle$ and $|\psi_4\rangle$ have distinct orbital character, it is necessary to apply a combination of a transverse magnetic field and strain to couple to the spin qubit. We next consider possible scenarios that allow the transition $|\psi_1\rangle \leftrightarrow|\psi_4\rangle$ including (i) the application of a quasi-static magnetic field, (ii) a microwave drive, and (iii) an optical Raman scheme. 

\subsection{Quasi-static magnetic field}
To allow the spin-qubit states to couple to strain we add to the system a perturbation in the form of an $x-$polarized magnetic field:
\begin{align}
H_{ B_x}=\left(
    \begin{array}{cccc}
     0 & 0 & B_x \gamma _{\rm S} & 0 \\
     0 & 0 & 0 & B_x \gamma _{\rm S} \\
     B_x \gamma _{\rm S} & 0 & 0 & 0 \\
     0 & B_x \gamma _{\rm S} & 0 & 0 \\
    \end{array}.
\right)
\end{align}
In the lowest order of perturbation theory, this Hamiltonian causes the following modification to the system eigenstates:
\begin{align}
    |\psi'_1\rangle&\approx|\psi_1\rangle+\frac{B_x \gamma _{\rm S}}{\nu_1-\nu_3}|\psi_3\rangle,\label{eq:psiprime1}\\
    |\psi'_2\rangle&\approx|\psi_2\rangle+\frac{B_x \gamma _{\rm S}}{\nu_2-\nu_4}|\psi_4\rangle,\\
    |\psi'_3\rangle&\approx|\psi_3\rangle+\frac{B_x \gamma _{\rm S}}{\nu_3-\nu_1}|\psi_1\rangle,\\
    |\psi'_4\rangle&\approx|\psi_4\rangle+\frac{B_x \gamma _{\rm S}}{\nu_4-\nu_2}|\psi_2\rangle.\label{eq:psiprime4}
\end{align}
The two lowest-lying spin states $|\psi_4\rangle$ and $|\psi_1\rangle$ are therefore modified to $|\psi'_4\rangle$ and $|\psi'_1\rangle$ which can be coupled via strain. In particular, in the lowest order of perturbation theory, this coupling can be estimated as: 
\begin{align}
    \langle \psi'_{4}| H_{\beta} |\psi'_1\rangle &\approx \beta\left( \frac{B_x \gamma _{\rm S}}{\nu_1-\nu_3}+\frac{B_x \gamma _{\rm S}}{\nu_4-\nu_2} \right)\nonumber\\
    &=\beta\left( \frac{B_x \gamma _{\rm S}}{-2B_z\gamma_{\rm S}-2\lambda}+\frac{B_x \gamma _{\rm S}}{2B_z\gamma_{\rm S}-2\lambda} \right).
\end{align}
Similarly for the $\gamma$ component of strain we obtain:
\begin{align}
    \langle \psi'_{4}| H_{\gamma} |\psi'_1\rangle &\approx -{\rm i}\gamma\left( \frac{B_x \gamma _{\rm S}}{\nu_1-\nu_3}+\frac{B_x \gamma _{\rm S}}{\nu_4-\nu_2} \right)\nonumber\\
    &=-{\rm i}\gamma\left( \frac{B_x \gamma _{\rm S}}{-2B_z\gamma_{\rm S}-2\lambda}+\frac{B_x \gamma _{\rm S}}{2B_z\gamma_{\rm S}-2\lambda} \right).
\end{align}
Based on our simulations we further consider $\beta/(2\pi)\sim 10$\,MHz ($\gamma/(2\pi)\sim 10$\,MHz) we obtain that the direct coupling of states $|\psi'_1\rangle$ and $|\psi'_4\rangle$, $g_{\rm pe}=\Gamma_{\rm pe}B_x$ is going to be of the order $\Gamma_{\rm pe}/(2\pi)\sim  5$\,MHz/T. A moderate magnetic bias field of 0.2 T would therefore be required to achieve the coupling rate $g_{\rm pe}/(2\pi)\sim 1$\,MHz used in the state-transfer analysis. The frequency of the spin transition $|\psi_{1}'\rangle\leftrightarrow |\psi_{4}'\rangle$ can be tuned by an external field $B_{z}$ to achieve resonant spin-phonon interaction. The pulsed modulation of the coupling could be realized by modulating the value of the magnetic field $B_x(t)$.

\subsection{Microwave drive}
Another way to induce the resonant interaction of the lowest lying spin states (states $|\psi_1\rangle$ and $|\psi_4\rangle$) considering that $\lambda$ is the dominant scale is to drive the spin transition between states $|\psi_4\rangle$ and $|\psi_1\rangle$ that are orbitally allowed via a microwave drive at the correct frequency $\omega_{\rm d}$ (that we determine later) as shown in \cite{lemonde2018phononnetworks}. This scheme is schematically depicted in Fig.\,\ref{fig:strainspin}(c). The orbital transitions $|\psi_1\rangle \to |\psi_2\rangle$ and $|\psi_4\rangle \to |\psi_3\rangle$ are coupled to the acoustic phonon via the strain susceptibility with a rate $g_{\rm orb}\approx 2\pi\times 10$\,MHz. We further introduce the shorthand notation: $\sigma_{ij}=|\psi_i\rangle\langle\psi_j |$ and write the effective Hamiltonian of the system under consideration:
\begin{align}
    H_{\rm sys}&=\Delta \sigma_{22}+\omega_{\rm B}\sigma_{44}+\Omega(t)\left( e^{{\rm i}[\theta(t)+\omega_{\rm d}t]}\sigma_{42}+{\rm H.c.}\right)\nonumber\\
    &+g_{\rm orb}\left (b^\dagger\sigma_{12}+{\rm H.c.}\right)+\omega_{\rm p} b^\dagger b.\label{eq:ham4level}
\end{align}
Here we neglected any influence of the off-resonant state $|\psi_{3}\rangle$, $\Delta=E_2-E_1$, $\omega_{\rm B}=E_4-E_1$, $\omega_{\rm p}$ is the phonon frequency, $b$ ($b^{\dagger}$) is the phonon annihilation (creation) operator, and $\Omega(t)$ and $\theta(t)$ are the amplitude- and phase-envelope of the external microwave drive, respectively. The Hamiltonian in Eq.\,\eqref{eq:ham4level} can be used to find an approximation in such a way that the Raman-mediated coupling of the two lowest spin states with the phonon can be made explicit. To that end we first introduce the interaction picture given by the Hamiltonian:
\begin{align}
    H_{\rm ip}=\omega_{\rm B}\sigma_{44}+\Delta\sigma_{22}+\omega_{\rm p} b^\dagger b.
\end{align}
This leads to the following rotating-frame Hamiltonian:
\begin{align}
    H_{\rm rot}&=\Omega(t)\left[ \sigma_{42}e^{{\rm i}[\theta(t)+(\omega_{\rm d}+\omega_{\rm B}-\Delta)]t}+{\rm H.c.}\right]\nonumber \\
    &+g_{\rm orb}\left [b^\dagger\sigma_{12}e^{{\rm i}(\omega_{\rm p}-\Delta)t}+{\rm H.c.}\right].\label{eq:mwdriveham}
\end{align}
We further set $\omega_{\rm d}=\omega_{\rm p}-\omega_{\rm B}$ to ensure resonant drive. 

Adiabatic elimination can be applied to the Hamiltonian in Eq.\,\eqref{eq:mwdriveham} to obtain the effective coupling constant between the phonon and the electron spin:
\begin{align}
    g_{\rm p-e}\approx g_{\rm eff}^{\rm mw}= g_{\rm orb}\frac{\Omega (t)e^{{\rm i}\theta(t)}}{\delta},
\end{align}
with $\delta=\omega_{\rm p}-\Delta$.
To ensure the validity of the adiabatic approximation we further require that $|\Omega|< |\delta|$ and we therefore estimate $g_{\rm pe}/(2\pi)=g_{\rm eff}^{\rm mw}/(2\pi)\approx 0.1 g_{\rm orb}/(2\pi)\approx 1$\,MHz. The microwave drive employed in this scheme ensures the resonant character of the phonon-spin coupling and eliminates the necessity to tune the magnitude of the magnetic field $B_z$ (i.e. of $\omega_{\rm B}$).  
%In conclusion, the coupling strength of an acoustic wave with the ground-state spin of SiV$^-$ is going to be approximately of the order of $2\pi\times 1$\,MHz.

\subsection{Optical Raman drive}
Finally, an optical Raman drive has been proposed \cite{lemonde2018phononnetworks} to enable resonant coupling between the transition connecting the lowest-energy spin-orbit states and the cavity phonon, as shown in Fig.\,\ref{fig:strainspin}(d). 
%Here we brifly summarize how the effective resonant coupling of the phonon to the electron spin transition $|\psi'_1\rangle$ and $|\psi'_4\rangle$ can be achieved. 
The Hamiltonian describing this Raman scheme, expressed in the basis of states perturbed by the magnetic field [Eqs.\,\eqref{eq:psiprime1} to \eqref{eq:psiprime4}] and an optically accessible excited state, $|\rm E\uparrow\rangle$, can be written as:
\begin{align}
    H_{\rm Raman}&=\Delta\sigma'_{22}+\omega_{\rm B}\sigma'_{44}+\omega_{\rm p} b^\dagger b+\omega_{\rm E}\sigma'_{\rm EE}\nonumber\\
    &+\Omega_{\rm A}(\sigma'_{\rm 2E}e^{{\rm i}[\theta_{\rm A}(t)+\omega_{\rm A}t]}+{\rm H.c.})\nonumber\\
    &+\Omega_{\rm C}(\sigma'_{\rm 4E}e^{{\rm i}[\theta_{\rm C}(t)+\omega_{\rm C}t]}+{\rm H.c.})\nonumber\\
    &+g_{\rm orb}(\sigma'_{12}b^\dagger + \sigma_{21}b),
\end{align}
where $\sigma'_{ij}=|\psi'_{i}\rangle\langle \psi'_{j}|$ and $|\psi'_{\rm E}\rangle$ is an electronic excited state of SiV$^{-}$. Here $\Omega_{\rm A}$ and $\Omega_{\rm B}$ are related to the amplitude of the two pumping lasers and are proportional to the dipole coupling elements between the respective states, and $\theta_{\rm A}$ ($\theta_{\rm C}$) are slowly varying phases. The respective laser-drive frequencies $\omega_{\rm A}$ and $\omega_{\rm C}$ are adjusted so that $\omega_{\rm p}=\omega_{\rm B}+\omega_{\rm A}-\omega_{\rm C}$. Under such conditions, it is possible to obtain the following effective phonon-electron-spin coupling:
\begin{align}
    g_{\rm p-e}\approx g_{\rm eff}^{\rm Raman}= \frac{\Omega_{\rm A}e^{{\rm i}\theta_{\rm A}(t)}\Omega_{\rm C}e^{-{\rm i}\theta_{\rm C}(t)}g_{\rm orb}}{(\omega_{\rm p}-\Delta)(\omega_{\rm C}-\omega_{\rm E}+\omega_{\rm p})}.
\end{align}
Since $g_{\rm eff}^{\rm Raman}$ has been obtained perturbatively, it is necessary that $\Omega_{\rm A}\Omega_{\rm C}/[(\omega_{\rm p}-\Delta)(\omega_{\rm C}-\omega_{\rm E}+\omega_{\rm p})]\ll 1$, and the effective phonon-electron-spin coupling is thus substantially reduced. The advantage of this scheme is in the tunnability of the externally applied lasers that can be used to rapidly adjust the condition for the resonant phonon-spin coupling or modulate the magnitude of the coupling strength. Notice also that in order for this scheme to be efficient, the phonon frequency must be close to the transition frequency $\Delta$. %This scheme is therefore more suitable to couple higher-frequency photons to the spin transition and is not suitable in the situation where $\omega_{\rm p}\ll\Delta$ considered here.

Last we mention that the different strain susceptibility of the ground and excited electronic-state manifolds could also be used to induce the ground-state spin-strain coupling. This scheme has, for example, been described in Ref. \cite{kuzyk2018scaling} for a nitrogen-vacancy color center.   

\section{Coordinate transformation of the strain tensor components}\label{app:transformed}

In Appendix\,\ref{app:strain} we discuss the effects of strain on the fine-structure states of a SiV$^-$ color center and express the strain tensor in the internal system of coordinates of the color center defined with respect to the diamond crystallographic directions as: $z$ along $[111]$,  $x$ along $[\bar{1}\bar{1}2]$, and $y$ along $[\bar{1}10]$. However, in applications it is more natural to consider the strain tensor in the set of coordinates defined by the basis vectors of the diamond cubic lattice. For convenience, we therefore transform the relevant tensor components that yield electron-spin-phonon coupling into this natural coordinate system defined by the basis vectors of the diamond cubic lattice and use the numbered indexes 1, 2, and 3 to denote the coordinates $[100]$, $[010]$, and $[001]$, respectively:
\begin{align}
\epsilon_{xx}-\epsilon_{yy} &=(-\epsilon _{11}-\epsilon _{22}+2 \epsilon _{33}+2 [\epsilon _{12}+ \epsilon _{21}]\nonumber\\
&-[\epsilon _{13}+\epsilon_{31}]-[\epsilon _{23}+\epsilon _{32}])/3\\
\epsilon_{zx}&=-(\epsilon _{11}+\epsilon _{22}-2 \epsilon _{33}-2\epsilon _{13}-2\epsilon _{23}\nonumber\\
&+\epsilon_{12}+\epsilon _{21}+\epsilon _{31}+\epsilon _{32})/(3\sqrt{2})\\
\epsilon_{xy}&=\frac{\epsilon _{11}-\epsilon _{12}+\epsilon _{21}-\epsilon _{22}-2 \epsilon _{31}+2 \epsilon _{32}}{2 \sqrt{3}}\\
\epsilon_{yz}&=\frac{-\epsilon _{11}-\epsilon _{12}-\epsilon _{13}+\epsilon _{21}+\epsilon _{22}+\epsilon _{23}}{\sqrt{6}}.
\end{align}
This form is convenient to express the effect of strained diamond slab etched along the $(100)$ crystalographic plane of diamond, which we consider in the design of the phononic cavity. 

\section{State transfer from the electron spin to the nuclear spin}\label{app:nucspin}
To complete the chain of state-transfer steps leading to the transduction of a state stored in an SC qubit to a nuclear-spin qubit, following reference \cite{bradley2019register} we discuss an example of a state-transfer protocol than can be applied to perform the step connecting the electronic and nuclear-spin qubits.  

We assume that the nuclear spin described by the Hamiltonian 
\begin{align}
H_{\rm nn}=\frac{\hbar\omega_{\rm L}}{2}\sigma_{z}^{\rm n}
\end{align}
is coupled to the electron spin via a longitudinal interaction:
\begin{align}
    H_{\rm e-n}=\frac{A_\parallel}{4} \sigma^{\rm e}_{z}\sigma^{\rm n}_z,
\end{align}
where $\sigma_{z}^{\rm e}$ and $\sigma_{z}^{\rm n}$ are the electron-spin and nuclear-spin Pauli $z$ operators, respectively. This interaction Hamiltonian is a result of a hyperfine interaction between the electronic and the nuclear spin. The nuclear spin is furthermore driven by a microwave field of frequency $\omega_{\rm mw}=\omega_{\rm L}+\frac{A_{\parallel}}{2}$, amplitude $\Omega_{\rm mw}$, and adjustable phase $\theta_{\rm mw}$:
\begin{align}
    H_{\rm mw}=\Omega_{\rm mw}\left[\sigma_{\rm n}e^{{\rm i}(\theta_{\rm mw}+\omega_{\rm mw}t)}+\sigma^\dagger_{\rm n}e^{-{\rm i}(\theta_{\rm mw}+\omega_{\rm mw}t)}\right].
\end{align}
This drive is conditionally resonant when the electron spin is in state $|1_{\rm e}\rangle$ and is off-resonant when the electron is in $|0_{\rm e}\rangle$. After transforming the total Hamiltonian $H_{\rm nn}+H_{\rm e-n}+H_{\rm mw}$ into an interaction picture and considering the conditional character of the drive, we obtain the effective Hamiltonian $H_{\rm en}$:
\begin{align}
    H_{\rm en}&=-\hbar {A_{\parallel}}\sigma^\dagger_{\rm n}\sigma_{\rm n}|0_{\rm e}\rangle\langle 0_{\rm e}|\nonumber\\
    &+\hbar\Omega_{\rm mw}\left[\cos(\theta_{\rm mw})\sigma^{\rm n}_x+\sin(\theta_{\rm mw})\sigma^{\rm n}_y\right]|1_{\rm e}\rangle\langle 1_{\rm e}|.\label{eq:hamen}
\end{align}
Importantly, $H_{\rm en}$ describes the time evolution of the system accurately only if $\Omega_{\rm mw}\ll A_{\parallel}$. When the electron spin is in $|0_{\rm e}\rangle$ the nuclear spin undergoes a free precession with an angular velocity $-A_{\parallel}$, and when the electron spin is in $|1_{\rm e}\rangle$ the nuclear spin rotates around an axis ${\bf e}_{\theta_{\rm mw}}=\cos(\theta_{\rm mw}){\bf e}_x+\sin(\theta_{\rm mw}){\bf e}_y$ (${\bf e}_x, {\bf e}_y$ being unit vectors along $x$ and $y$, respectively) with angular velocity $2\Omega_{\rm mw}$. 

We next consider that the electron spin is periodically flipped via a dynamical decoupling sequence of the form $(\tau - \pi - 2\tau - \pi - \tau)^{N/2}$, where $N$ is an (even) number of pulses applied to the system. The total duration of the pulse sequence is $T_{N}=2N\tau$ and we consider that the gate applied to the nuclear spin is completed at $t=T_{N}$. The phase $\theta_{\rm mw}$ of the microwave drive must be adjusted after each pulse $k$ as:
\begin{align}
    \theta_{\rm mw}&=(k-1)\phi_{k}+\phi_{c}+\phi_0 & {\rm for}\; k\;{\rm odd},\nonumber\\
\theta_{\rm mw}&=(k-1)\phi_{k}+\phi_0 & {\rm for}\; k\;{\rm even},\nonumber\\
\end{align}
where $\phi_k=-(2-\delta_{1k})\tau A_{\parallel}$, and $\phi_c=0$ for unconditional rotations of the nuclear spin ($\phi_c=\pi$ for conditional rotations of the nuclear spin). The angle of rotation $\varphi$ of the nuclear spin about the axis determined by $\cos(\phi_0){\bf e}_x+\sin(\phi_0){\bf e}_y$ is $\varphi=2\Omega_{\rm mw}\tau N$. The Rabi frequency $\Omega_{\rm mw}$ must therefore be appropriately adjusted in order to achieve the desired rotation angle $\varphi$. We denote the unconditional gate implemented by the above described protocol as $R_{\phi_0, \varphi}^{\rm n}$ and the conditional gate as $C_{\phi_0, \varphi}^{\rm n}$. The conditional gate rotates the nuclear spin by an angle $-\varphi$ if the electron spin is initially in $|1_{\rm e}\rangle$. The following sequence of controlled and uncontrolled rotations produces a SWAP gate exchanging the states of the electron and the nuclear spin:
%The interaction Hamiltonian $H_{\rm e-n}$ can be further simplified by rewriting $\sigma_z^{\rm e}=2\sigma_{\rm e}^\dagger\sigma_{\rm e}-I_{\rm e}$ ($\sigma_z^{\rm n}=2\sigma_{\rm n}^\dagger\sigma_{\rm n}-I_{\rm n}$), which produces:
%\begin{align}
%     H_{\rm e-n}=\hbar A_{\rm ne}\sigma_{\rm e}^\dagger\sigma_{\rm e}\sigma_{\rm n}^\dagger\sigma_{\rm n}-\frac{\hbar A_{\rm ne}}{2}(\sigma_{\rm e}^\dagger\sigma_{\rm e}+\sigma_{\rm n}^\dagger\sigma_{\rm n})+\frac{\hbar A_{\rm ne}}{4}.\label{eq:hamentr}
%\end{align}
%The first term in Eq.\,\eqref{eq:hamentr} represents a conditional phase shift of the state in which both the electronic and the nuclear spin are excited ($|1_{\rm e}1_{\rm n}\rangle$). We denote this phase gate $C_\phi$, where $\phi$ is the additional phase acquired by state $|1_{\rm e}1_{\rm n}\rangle$. The second term in Eq.\,\eqref{eq:hamentr} can be absorbed into the definition of energies of the respective qubits and the last term is a constant that does not influence system dynamics. We further assume that each spin can be individually addressed via MW-pulse irradiation which allows for implementation of arbitrary spin rotations (single-qubit gates). Assuming that the system is initially in state $|\psi_{\rm i}\rangle=\alpha|0_{\rm e}0_{\rm n}\rangle+\beta|1_{\rm e}0_{\rm n}\rangle$, the following sequence of single- and two-qubit gates can be applied to accomplish the state transfer to $|\psi_{\rm f}\rangle=-\alpha|0_{\rm e}0_{\rm n}\rangle-\beta|0_{\rm e}1_{\rm n}\rangle$:
\begin{align}
    |\psi_{\rm f}\rangle =CX^{\rm n}\cdot H^{\rm e}\cdot H^{\rm n}\cdot CX^{\rm n}\cdot H^{\rm e}\cdot H^{\rm n}\cdot CX^{\rm n}|\psi_{\rm i}\rangle,
\end{align}
where $CX^{\rm n}$ is the controlled not gate conditionally flipping the nuclear spin, $H^{\rm s}$ is the single-qubit Hadamard gate acting on the electron qubit, ${\rm s}={\rm e}$, or the nuclear qubit, ${\rm s}={\rm n}$. The single and two-qubit gates outlined above can be constructed from the conditional rotation of the nuclear spin and local qubit operations. In particular, the Hadamard gate acting on the nuclear spin can be constructed as $H^{\rm n}=R^{\rm n}_{0, \pi}\cdot R^{\rm n}_{\frac{\pi}{2}, \frac{\pi}{2}}$. Similarly, $CX^{\rm n}=S_{\frac{\pi}{2}}\cdot R^{\rm n}_{\rm 0, \frac{\pi}{2}}\cdot C^{\rm n}_{\rm 0, \frac{\pi}{2}}$, with $S_{\frac{\pi}{2}}=\sigma_{\rm e}\sigma^\dagger_{\rm e}+{\rm i}\sigma^\dagger_{\rm e}\sigma_{\rm e}$ (a rotation around the $z$ axis).

Note that the time-duration of the single-qubit rotations applied to the electron spin is mainly dependent on the intensity of the applied pulses and we treat it as practically instantaneous. On the other hand, the gates applied to the nuclear spin rely on a free time evolution of the system limited by $\Omega_{\rm mw}\ll A_{\parallel}$. This sets the limit to the achievable state-transfer fidelity $\mathcal{F}_{\rm en}$ when spin dephasing is taken into account. We phenomenologically account for pure dephasing of both the electron and the nuclear spin via the Lindblad superoperators $\gamma_{\rm e}\mathcal{L}_{\sigma^\dagger_{\rm e}\sigma_{\rm e}}(\rho)$ and $\gamma_{\rm n}\mathcal{L}_{\sigma^\dagger_{\rm n}\sigma_{\rm n}}(\rho)$ [see Eq.\,\eqref{eq:lindblad}] that together with $H_{\rm en}$ [Eq.\,\eqref{eq:hamen}] describe the dynamics of the system. We estimate the fidelity of the state transfer performed by the SWAP gate for a moderate value of the longitudinal spin-spin coupling $A_{\parallel}/(2 \pi)= 500$\,\si{\kilo\hertz} and we set the drive frequency to $\Omega_{\rm mw}/(2 \pi)\approx 3.9$\,\si{\kilo\hertz}. We further consider $\gamma_{\rm e}/(2 \pi)= 10$\,\si{\kilo\hertz} and $\gamma_{\rm n}/(2 \pi) = 1$\,\si{\hertz}. With these values we estimate $\mathcal{F}_{\rm en}\approx 0.9975$, as given in the main text. 

\section{Two-qubit gates applicable to the electron spins}\label{app:molmersor}
\begin{figure}[t!]
    \centering
    \includegraphics[scale=1]{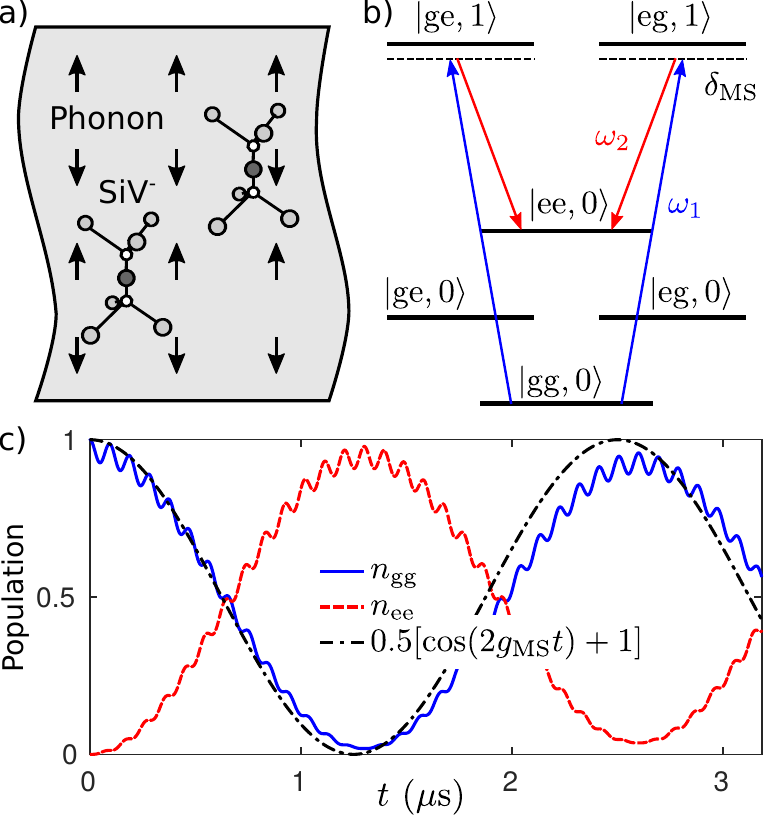}
    \caption{M{\o}lmer-S{\o}rensen gate can be used to entangle electron spins. (a) Two electron spins of diamond color centers interact with a shared mechanical mode of the phononic cavity. (b) A two-photon drive can be used to perform a transition from the ground state of the system $|{\rm gg}, 0\rangle$ to the doubly excited state $|{\rm ee},0\rangle$ without changing the number of phonons in the system or populating an intermediate state. (c) Dynamics of the ground-state ($|{\rm gg}, 0\rangle$) population, $n_{\rm gg}$, and the doubly excited-state ($|{\rm ee},0\rangle$) population $n_{\rm ee}$ compared with the ideal time dependence $0.5[\cos(2g_{\rm MS}t)+1]$ following from the effective model implementing the M{\o}lmer-S{\o}rensen gate.}\label{fig:molmersorensen}
\end{figure}
One of the advantages of the architecture proposed in this paper is that the color-center electron spins can be used to prepare non-classical many-body quantum-mechanical states that can be further utilized for processing of quantum information, quantum teleportation, or speedup of quantum-state transduction.  
In this appendix we provide a suggestion of a gate that could be used to generate a GHZ state (i.e. an entangled Bell state) of a pair of electron-spin qubits coupled to a common vibrational mode. In Appendix\,\ref{app:strain} we have shown that the electron-spin states can be coupled to a strain field via effective controllable coupling schemes. This leads to the effective interaction between a mode of an acoustical cavity coupled to two electron spins:
\begin{align}
    H_{\rm eff}&=\hbar\omega_{\rm e1}\sigma_{\rm e1}^\dagger \sigma_{\rm e1}+\hbar\omega_{\rm e2}\sigma_{\rm e2}^\dagger \sigma_{\rm e2}+ \hbar\omega_{\rm p} b^\dagger b \nonumber\\
    & +\hbar g_{\rm eff}(\sigma_{\rm e1}+\sigma_{\rm e1}^\dagger) (b+b^\dagger)\nonumber\\
    &+\hbar g_{\rm eff}(\sigma_{\rm e2}+\sigma_{\rm e2}^\dagger) (b+b^\dagger).\label{eq:gateham1}
\end{align}
Here $\sigma_{\rm e1}=| 0_1\rangle\langle 1_1 |$ ($\sigma_{\rm e2}=| 0_2\rangle\langle 1_2 |$) are the lowering operators of the respective two-level spin systems, $b$ ($b^\dagger$) is the annihilation (creation) operator of the shared phonon mode, $\omega_{\rm e1}$ and $\omega_{\rm e2}$ are the frequencies of the respective spins, and $\omega_{\rm p}$ is the frequency of the phonon mode. The effective coupling $g_{\rm eff}$ can be realized as described in Appendix\,\ref{app:strain}. 
It is more convenient to transform the Hamiltonian in Eq.\,\eqref{eq:gateham1} into the interaction picture:
\begin{align}
    H_{\rm eff}^{\rm I}&=g_{\rm eff}(\sigma_{\rm e1}e^{-{\rm i}\omega_{\rm e1}t}+\sigma_{\rm e1}^\dagger e^{{\rm i}\omega_{\rm e1}t}) (b e^{-{\rm i}\omega_{\rm p}t}+b^\dagger e^{{\rm i}\omega_{\rm p}t})\nonumber\\
    &+g_{\rm eff}(\sigma_{\rm e2}e^{-{\rm i}\omega_{\rm e2}t}+\sigma_{\rm e2}^\dagger e^{{\rm i}\omega_{\rm e2}t}) (b e^{-{\rm i}\omega_{\rm p}t}+b^\dagger e^{{\rm i}\omega_{\rm p}t}).\label{eq:gateham1INT}
\end{align}
Next, we assume that the coupling $g_{\rm eff}$ can be modulated in time as $g_{\rm eff}(t)=\frac{g_{\rm eff}^0}{4}(e^{{\rm i}\omega_1 t}+e^{{\rm i}\omega_2 t}+{\rm H.c.})$ (H.c. stands for the Hermitian conjugate). We assume a situation where $\omega_{\rm p}> \omega_{\rm e}=\omega_{\rm e1}=\omega_{\rm e2}$ and therefore select the two drive frequencies as $\omega_1=\omega_{\rm e}+\omega_{\rm p}-\delta_{\rm MS}$, and $\omega_2=\omega_{\rm p}-\omega_{\rm e}-\delta_{\rm MS}$, with $\delta_{\rm MS}$ a small detunning. We further simplify the Hamiltonian by assuming $\omega_{\rm e1}=\omega_{\rm e2}=\omega_{\rm e}$. The interaction-picture Hamiltonian [Eq.\,\eqref{eq:gateham1INT}] then becomes (considering only slowly oscillating terms in the RWA):
\begin{align}
    H_{\rm eff}^{\rm RWA}&\approx g^0_{\rm eff}[(\sigma_{\rm e1}+\sigma_{\rm e2})b^\dagger e^{-{\rm i}(\omega_{\rm e}-\omega_{\rm p}+\omega_{2})t}\nonumber\\
    &+ (\sigma_{\rm e1}^\dagger +\sigma_{\rm e2}^\dagger)b^\dagger e^{{\rm i}(\omega_{\rm e}+\omega_{\rm p}-\omega_{1})t}+{\rm H.c.}]. \label{eq:gatehamdrivenINT}
\end{align}
From this Hamiltonian we can obtain the effective coupling $g_{\rm M-S}$ between the state $|{0_1}\rangle\otimes|{0_2}\rangle\otimes |0 \rangle\equiv|{\rm gg},0\rangle$ and the doubly excited state $|1_1\rangle\otimes|{1_2}\rangle\otimes |0 \rangle\equiv|{\rm ee},0\rangle$ (more generally $|{\rm gg},n\rangle$ and $|{\rm ee},n\rangle$, with $n$ the number of phonons):
\begin{align}
    g_{\rm M-S}\approx \frac{(g_{\rm eff}^0)^2}{8\delta_{\rm MS}}.
\end{align}
We plot the resulting dynamics of the populations of the two excited states in Fig.\,\ref{fig:molmersorensen}. The population of the state $|{\rm gg}, 0\rangle$ (blue line) coherently transfers into the population of $|{\rm ee},0\rangle$ (orange line). For comparison we plot the expression $0.5[\cos(2g_{\rm MS}t)+1]$ as the yellow line in Fig.\,\ref{fig:molmersorensen}. We consider that both electron spins and the phonon are subject to decoherence as described in the main text. If we stop the time evolution at $t\approx 0.675$\,\si{\micro\second} we obtain a highly entangled Bell state, the two-qubit GHZ state (up to a phase factor).

\bibliography{biblio}
\end{document}